\documentclass[prd,10pt,aps,twocolumn,nofootinbib,showpacs,floatfix,amssymb,amsmath]{revtex4-1}
\usepackage{slashed,rotating,esint,hyperref}

\newcommand{\ch}{\mathrm{ch}}

\newcommand{\beqn}{\begin{eqnarray}}
\newcommand{\eeqn}{\end{eqnarray}}
\newcommand{\eq}[1]{(\ref{#1})}
\newcommand{\bl}{{\biggl|}}

\newcommand{\cL}{{\cal L}}

\newcommand{\sign}{{\mathrm{sgn}}\,}

\newcommand{\Z}{{\mathbb{Z}}}

\begin{document}

\title{Rotating Casimir systems: magnetic--field--enhanced perpetual motion, possible realization in doped nanotubes, and laws of thermodynamics}
\author{M. N. Chernodub}\email{maxim.chernodub@lmpt.univ-tours.fr; \\ On leave from ITEP, Moscow, Russia.}
\affiliation{CNRS, Laboratoire de Math\'ematiques et Physique Th\'eorique, Universit\'e Fran\c{c}ois-Rabelais Tours,\\ F\'ed\'eration Denis Poisson, Parc de Grandmont, 37200 Tours, France}
\affiliation{Department of Physics and Astronomy, University of Gent, Krijgslaan 281, S9, B-9000 Gent, Belgium}

\begin{abstract}
Recently, we have demonstrated that for a certain class of Casimir--type systems (``devices'') the energy of zero--point vacuum fluctuations reaches its global minimum when the device rotates about a certain axis rather than remains static. This rotational vacuum effect may lead to the emergence of permanently rotating objects provided the negative rotational energy of zero--point fluctuations cancels the positive rotational energy of the device itself. In this paper, we show that for massless electrically charged particles the rotational vacuum effect should be drastically (astronomically) enhanced in the presence of a magnetic field.  As an illustration, we show that in a background of experimentally available magnetic fields the zero--point energy of massless excitations in rotating torus--shaped doped carbon nanotubes may indeed overwhelm the classical energy of rotation for certain angular frequencies so that the permanently rotating state is energetically favored. The suggested ``zero--point driven'' devices  -- which have no internally moving parts -- correspond to a perpetuum mobile of a new, fourth kind: They do not produce any work despite the fact that their equilibrium (ground) state corresponds to a permanent rotation even in the presence of an external environment. We show that our proposal is consistent with the laws of thermodynamics.
\end{abstract}


\date{August 24, 2012}

\maketitle

\section{Introduction}

The most direct manifestation of the existence of zero--point fluctuations of the vacuum is the Casimir effect~\cite{ref:Casimir}. The essence of the Casimir effect, in its original formulation, is the appearance of an attracting force between two parallel infinitely large plates in the vacuum. The plates are taken to be perfectly conducting and electrically neutral so that the attracting force appears only due to the  vacuum fluctuations of the electromagnetic field. 

The mechanism of the Casimir effect is as follows: The presence of the plates affects the zero--point fluctuations of the electromagnetic field thus modifying their energy spectrum. Because of the change in the energy spectrum, the total energy of the vacuum fluctuations in the presence of the plates is different from the total energy of the fluctuations in the absence of the plates. Both energies are infinitely large but their difference is a finite quantity which is called the Casimir energy or the zero--point energy. In this particular example the Casimir energy is a negative quantity, the absolute value of which increases rapidly as the distance between the plates gets smaller. As a result the plates are attracting to each other. The existence of the attracting force was confirmed experimentally~\cite{ref:Casimir:experimental}. The current progress in this rapidly evolving field is reviewed in various books~\cite{ref:Milton:Book,ref:Bordag:Book,ref:Mostepanenko:Book}.

The interesting property of the zero--point energy is that it has a mass. More precisely, in the external gravitational field the Casimir energy gravitates as required by the equivalence principle, so that  the gravitational and inertial masses, associated with the Casimir energy $E_{C}$ are both $m_{C} = E_{C}/c^{2}$~\cite{Fulling:2007xa}. This statement is valid regardless of the sign of the Casimir energy and therefore the negative Casimir energy corresponds to a negative inertial mass. 

A negative mass should have a negative moment of inertia. As a result, if a system possessing a negative zero--point energy is rotated, then the rotational energy corresponding to zero--point fluctuations should {\it decrease} as the angular frequency is increased. This effect, called the rotational vacuum effect, was indeed found in Ref.~\cite{ref:I}. The fact, that the moment of inertia of zero-point fluctuations can take negative values, was later confirmed by a different method in Ref.~\cite{Schaden:2012dp}.

Moreover, if the (positive--valued) classical  moment of inertia of the system is small enough then at certain nonzero frequency $\Omega_{0} \neq 0$ the negative rotational energy of the zero--point fluctuations may make the total rotational energy  $E$ of the system smaller compared to the total rotational energy in the static state, $E(\Omega_{0}) < E(0)$. In this case the ground state of the system should correspond to $\Omega \neq 0$ and the system should prefer to rotate forever in its ground state even in the presence of external environment such as a thermal bath. This ``zero--point driven'' perpetual motion is philosophically similar to ``time crystals'' proposed recently in both specific classical and semiclassical as well as quantum--mechanical systems~\cite{ref:Shapere:Wilczek} and to the suggested ``space--time crystal'' system of a permanent current of cold ions~\cite{ref:CI}.

However, the rotational vacuum effect is extremely small. The typical energy scales involved in the rotational vacuum effect are as small as the energy scales of the usual Casimir effect~\cite{ref:I}. The first goal of this paper is to show that the rotational vacuum effect for electrically charged massless particles should be strongly enhanced in the presence of the magnetic field so that it can probably be tested experimentally in metallic carbon nanotubes. The second goal of this paper is to demonstrate that the very existence of the zero--point driven perpetual motion of a macroscopic system does not violate the laws of thermodynamics.
 
\begin{figure}[!thb]
\begin{center}
\includegraphics[scale=0.3,clip=false]{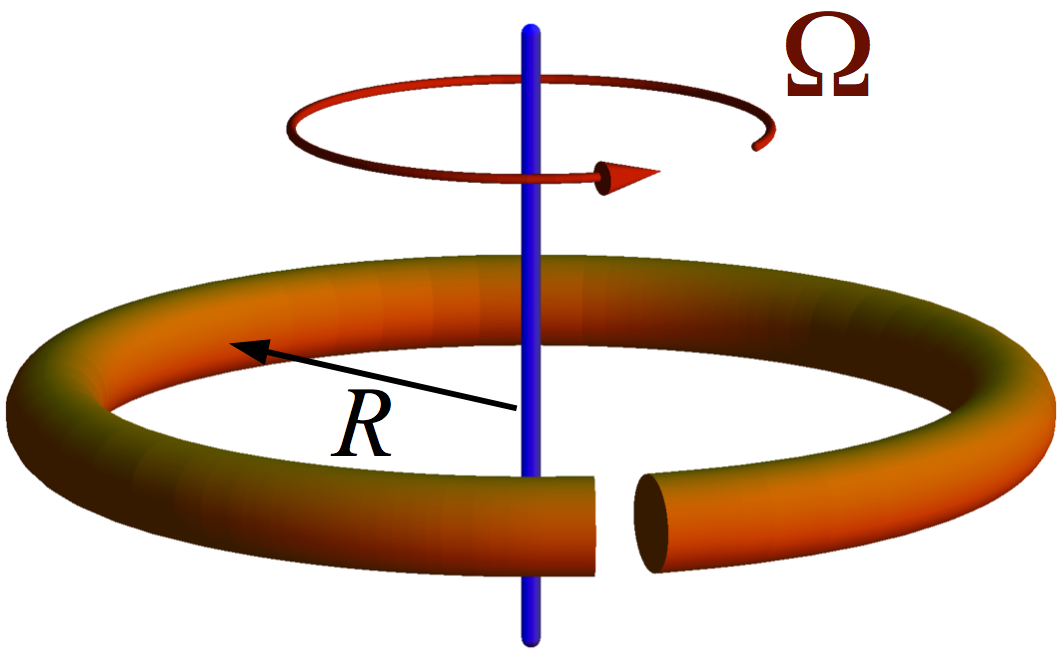}
\end{center}
\vskip -5mm
\caption{A simplest device which demonstrates the existence of the rotational vacuum effect (i.e., a negative moment of inertia of zero--point fluctuations): A massless scalar field is defined on a thin circle with an infinitesimally thin cut (an ``open ring''). The cut imposes a Dirichlet boundary condition on the field.}
\label{fig:setup}
\end{figure}

Our strategy is as follows. In Section~\ref{sec:neutral:1d} we provide details of the calculations outlined in Ref.~\cite{ref:I} and demonstrate that 
\begin{itemize}
\item[(i)] the rotational zero--point energy is negative; and 
\item[(ii)] the absolute value of the rotational zero--point energy increases with increase of angular frequency~$\Omega$. 
\end{itemize}
To this end we consider a simplest system exhibiting the rotational vacuum effect: a  massless scalar field defined on a thin circle with a point where the field vanishes (i.e., the point represents an infinitesimally thin ``Dirichlet cut''); see Fig.~\ref{fig:setup}. For shortness, we often call the infinitely thin circle with the Dirichlet cut as ``the device.''

In Section~\ref{sec:static:D} we consider a static device with a neutral scalar field. For methodological reasons, we overview the computation of the corresponding zero--point energy in a Green's function approach using both the time--splitting regularization and the $\zeta$--function regularization. For completeness, we also discuss a formal computation of the same zero-point energy via the $\zeta$--function regularization of a sum over the individual energies of all vacuum modes. Having reviewed these methods we find that it is very convenient to calculate the zero--point energy by using an explicit form of the Green's function and the time--splitting regularization. We will use the latter method henceforth.

In Section~\ref{sec:rotating:D} we repeat the calculation for a uniformly rotating circle with a cut. We demonstrate that the negative zero--point energy of the circle with the Dirichlet cut decreases further as the angular frequency of the rotation increases thus confirming our earlier calculation~\cite{ref:I}.

In Section~\ref{sec:charged:1d}  we consider the same device but with an electrically charged massless scalar field. We show that the background magnetic field drastically enhances the rotational vacuum effect. The enhancement depends quadratically on the number of elementary integer fluxes (quanta) of the magnetic field which pierce the circle. 

In Section~\ref{sec:nanotube} we propose to construct the device from a metallic carbon nanotube in a form of a closed torus with a cut made by a suitable chemical doping. We roughly estimate the rotational energy of the zero--point fluctuations in the device and demonstrate that the magnetic enhancement of the rotational vacuum effect may be reasonably large for experimental detection even in a very conservative estimation of the enhancement. As an example, we estimate the period $\tau$ of perpetual rotation (about one second) for the minimal radius $R$ of the torus (about the width of a human hair, 0.1~mm) at the highest experimentally available magnetic field ($B = 50$~Tesla). At weaker magnetic field the radius of the device and the time period should both increase.

In Section~\ref{sec:thermodynamics} we argue that the existence of such a device -- which can be considered as a ``perpetuum mobile of the fourth\footnote{We call it {\bf the perpetuum mobile of the fourth kind}, because the first three kinds do not exist, as they violate either the first or the second (or both) laws of thermodynamics.} kind driven by zero--point fluctuations'' -- is consistent with the laws of thermodynamics due to absence of the energy transfer and due to a discontinuous nature of the rotational energy of the zero--point fluctuations.

The last section is devoted to our conclusions.

\section{Example of spontaneous rotation: Real-valued scalar field on a circle with a Dirichlet cut}
\label{sec:neutral:1d}

\subsection{Nonrotating circle with the Dirichlet cut}
\label{sec:static:D}

\subsubsection{Lagrangian, boundary conditions and Casimir energy}

Consider a real-valued massless scalar field $\phi = \phi(t,\varphi)$ defined on a circle with a fixed radius $R$. The corresponding Lagrangian is:
\beqn
\cL & = & \frac{1}{2} \partial_{\mu} \phi \partial^{\mu} \phi \equiv  \frac{1}{2} \left[\left(\frac{\partial \phi}{\partial t}\right)^{2} - \frac{1}{R^{2}} \left(\frac{\partial \phi}{\partial \varphi}\right)^{2} \right] \,,
\label{eq:circle:L} 
\eeqn
where $\varphi \in [0, 2\pi)$ is the angular coordinate. 

In order to make the circle sensitive to rotations (this case will be considered in Section~\ref{sec:rotating:D}), we make an infinitesimally small cut at $\varphi = 0$; see Fig.~\ref{fig:setup}.
The role of the cut is to impose the simplest, Dirichlet boundary conditions on the field $\phi$ at the position of the cut:
\beqn
\phi (t,\varphi)\bl_{\varphi = 0} \equiv \phi (t,\varphi)\bl_{\varphi = 2 \pi}  = 0\,,
\label{eq:cirle:D:0}
\eeqn
where the points $\varphi = 0$ and $\varphi = 2 \pi$ are identified. For shortness, we call this cut as the Dirichlet cut. 

The field $\phi$ experiences quantum (zero--point) fluctuations. The mean energy density of the field fluctuations ${\bar {\mathcal E}}$ 
is given by a sum over energies $\varepsilon_{m}(R)$ of all individual fluctuation modes $m$:
\beqn
{\bar {\mathcal E}}(R) = \frac{1}{2 \pi R} \sum_{{\mathrm{modes}}\, m} \varepsilon_{m}(R)\,.
\label{eq:calE:R}
\eeqn 

The infinite sum in Eq.~\eq{eq:calE:R} is a divergent quantity both for the circle with a finite radius $R$ and in a free space ($R \to \infty$). However, the difference between these vacuum energy densities,
\beqn
{\bar {\mathcal E}}^{\mathrm{phys}}(R) = {\bar {\mathcal E}}(R) - {\bar {\mathcal E}}(\infty)\,,
\eeqn
is a finite physical quantity since the divergency in the energy density~\eq{eq:calE:R} is independent of the radius $R$. The total energy of the quantum fluctuations, 
\beqn
E^{\mathrm{phys}}(R) \equiv 2 \pi R\,  {\bar {\mathcal E}}^{\mathrm{phys}}(R)\,,
\label{eq:Casimir:energy}
\eeqn
is an experimentally measurable finite observable called the Casimir energy~\cite{ref:Casimir}.

\subsubsection{Energy of nonrotating circle: Time--splitting and $\zeta$--regularizations in Green's function approach}
\label{sec:static:Green:zeta}

The local energy density of the quantum fluctuations ${\mathcal E}(x)$ is given by the quantum expectation value of a certain component of the stress--energy tensor, $\left\langle T^{\mu\nu} (x) \right\rangle$:
\beqn
{\mathcal E}(x) = \left\langle T^{00} (x) \right\rangle\,.
\label{eq:E:T00}
\eeqn

The expectation value of the stress--energy tensor can be computed by using a Feynman--type Green function,
\beqn
G(x,x') = i \left\langle {\mathrm T} \phi (x) \phi (x') \right\rangle\,,
\label{eq:G:phi}
\eeqn
via the following relation~\cite{ref:Milton:Book,ref:Bordag:Book}:
\beqn
\left\langle T^{\mu\nu} (x) \right\rangle = \left( \partial^{\mu}\partial'^{\nu} - \frac{1}{2} g^{\mu\nu} \partial^{\lambda} \partial'_{\lambda}\right) \frac{1}{i} G(x,x') \bl_{x \to x'}\,. \qquad
\label{eq:Tmunu}
\eeqn
In Eq.~\eq{eq:G:phi} the symbol ``$\mathrm{T}$'' stands for the time--ordering operator.

In our case, the coordinates are $x \equiv (x^{0},x^{1}) = (t, R \varphi)$ and the corresponding derivatives are as follows:
\beqn
\partial_{0} = \frac{\partial}{\partial t}, \quad
\partial_{1} = 
\frac{1}{R} \frac{\partial}{\partial \varphi}, \quad
\partial_{0}' = \frac{\partial}{\partial t'}, \quad
\partial_{1}' = 
\frac{1}{R} \frac{\partial}{\partial \varphi'}. \qquad
\eeqn
The line element is:
\beqn
d s^{2} \equiv g_{\mu\nu} d x^{\mu} d x^{\nu} & = & (d x^{0})^{2} - (d x^{1})^{2} \nonumber \\
& \equiv & d t^{2} - R^{2} d \varphi^{2}, \quad
\eeqn
so that $g_{\mu\nu} = {\mathrm{diag}}\, (1, - 1)$ and
\beqn
\partial^{\lambda} \partial'_{\lambda} \equiv 
\frac{\partial}{\partial t} \frac{\partial}{\partial t'} - \frac{1}{R^{2}} \frac{\partial}{\partial \varphi} \frac{\partial}{\partial \varphi'}\,.
\eeqn

Then the energy density~\eq{eq:E:T00} is given by the following formula:
\beqn
\left\langle T^{00} (t,\varphi) \right\rangle & = & \left(\frac{\partial}{\partial t} \frac{\partial}{\partial t'} + \frac{1}{R^{2}}  \frac{\partial}{\partial \varphi} \frac{\partial}{\partial \varphi'}\right) 
\nonumber \\ & &  
\frac{1}{2 i} G(t,t';\varphi,\varphi') \bl_{{}^{t' \to t}_{\varphi' \to \varphi}}\,. \qquad
\label{eq:T00}
\eeqn

The Green function $G(t,t';\varphi,\varphi')$ in Eqs.~\eq{eq:G:phi} and \eq{eq:T00} satisfies the following equation:
\beqn
\left(\frac{\partial^{2}}{\partial t^{2}} {-} \frac{1}{R^{2}} \frac{\partial^{2}}{\partial \varphi^{2}}\right) G(t,\varphi;t',\varphi') {=} \frac{1}{R} \delta(\varphi - \varphi') \delta(t - t'), \qquad
\label{eq:Green:function}
\eeqn
which is valid in the region $0<\varphi, \varphi'<2 \pi$. Because of the Dirichlet boun\-da\-ry condition~\eq{eq:cirle:D:0}, the Green's function $G$ should vanish at the Dirichlet cut at $\varphi = 0,2\pi$ 
and at $\varphi' = 0,2\pi$.

It is convenient to express the Green function $G$ via the eigenvalues 
\beqn
\lambda_{\omega,m} = \frac{m^{2}}{4 R^{2}} - \omega^{2}\,, \qquad
\label{eq:eigenvalues}
\eeqn
and the eigenfunctions 
\beqn
\phi_{\omega,m}(t,\varphi) & \equiv & \phi^{(C)}_{\omega,m} (t,\varphi) - i \phi^{{(S)}}_{\omega,m} (t,\varphi) = e^{ - i \omega t} \, \chi_{m} (\varphi)\,, \qquad 
\label{eq:phi:omega}\\ 
\chi_{m} (\varphi)      &     =     & \frac{1}{\sqrt{\pi R}} \sin \frac{m \varphi}{2}
\label{eq:chi:static} 
\eeqn
of the second--order differential operator in the right hand side of Eq.~\eq{eq:Green:function}:
\beqn
\left(\frac{\partial^{2}}{\partial t^{2}} {-} \frac{1}{R^{2}} \frac{\partial^{2}}{\partial \varphi^{2}}\right) \phi_{\omega,m}(t,\varphi) {=} \lambda_{\omega,m} \, \phi_{\omega,m}(t,\varphi)\,, \qquad
\label{eq:phi:solutions}
\eeqn
where $m = 1,2,3,\dots$ and $\omega \in {\mathbb R}$. For convenience, the real-valued eigenfunctions $\phi^{(C)}_{\omega,m}$ and $\phi^{(S)}_{\omega,m}$ -- which have the same eigenvalues -- were combined into one complex eigenfunction~\eq{eq:phi:omega}. All eigenfunctions satisfy the Dirichlet boun\-da\-ry condition~\eq{eq:cirle:D:0} at the cut $\varphi = 0$.

The system of the eigenfunctions~\eq{eq:phi:omega} is orthonormal:
\beqn
&&  R \int\limits_{-\infty}^{+\infty} d t \int\limits_{0}^{2\pi} d \varphi \phi^{\dagger}_{\omega_{1},m_{1}}(t,\varphi)\phi_{\omega_{2},m_{2}}(t,\varphi) \nonumber \\
& & \hskip 25mm = 2 \pi \delta(\omega_{1} - \omega_{2}) \delta_{m_{1},m_{2}}\,,
\label{eq:orthonormal:static}
\eeqn
and complete:
\beqn
& & \int\limits_{-\infty}^{+\infty} \frac{d \omega}{2 \pi} \sum_{m=1}^{\infty} \phi_{m,\omega}(t_{1},\varphi_{1})  \phi^{\dagger}_{m,\omega}(t_{2},\varphi_{2}) 
=  \frac{1}{R} \, \delta \left( t_{1} - t_{2} \right) \nonumber \\
& & \qquad \cdot \sum_{n \in \Z} \bigl[\delta(\varphi_{1} - \varphi_{2} + 4 \pi n) - \delta(\varphi_{1} + \varphi_{2} + 4 \pi n) \bigr] \,.
\label{eq:complete:static}
\eeqn

Outside the boundary, $\varphi_{1,2} \neq 0,2\pi$, the right hand side of Eq.~\eq{eq:complete:static} is proportional to the product of the $\delta$ functions, 
$\delta(\varphi_{1} - \varphi_{2}) \delta(t_{1} - t_{2})$. At the boundary, the right hand side of Eq.~\eq{eq:complete:static} vanishes, as expected.

\begin{figure}[!thb]
\begin{center}
\includegraphics[scale=0.24,clip=false]{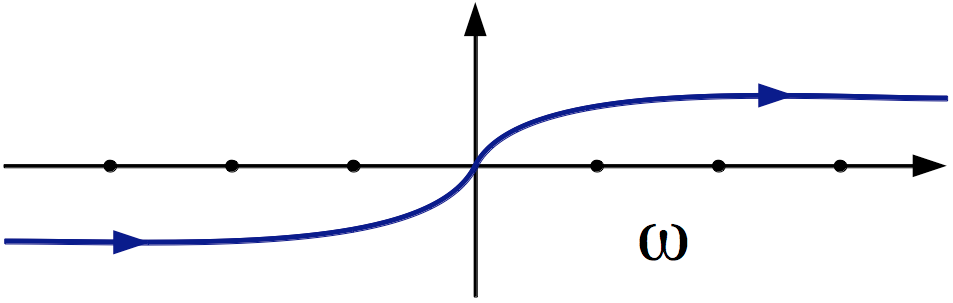}
\end{center}
\caption{The integration contour in the propagator~\eq{eq:G:general}.}
\label{fig:contour}
\end{figure}

The Green's function~\eq{eq:G:phi} is given by a general equation,
\beqn
G(t,t';\varphi,\varphi') & = & \int\limits_{-\infty}^{+\infty} \frac{d \omega}{2 \pi}  \sum_{m=1}^{\infty}
\frac{\phi_{\omega,m}(t,\varphi) \phi^{\dagger}_{\omega,m}(t',\varphi')}{\lambda_{\omega,m} - i \epsilon} \,, \qquad
\label{eq:G:general}
\eeqn
which in our case takes the following form:
\beqn
G(t,t';\varphi,\varphi') & = & \frac{1}{\pi R} \int\limits_{-\infty}^{+\infty} \frac{d \omega}{2 \pi}  \sum_{m=1}^{\infty} { \left(\frac{m^{2}}{4 R^{2}} - \omega^{2} - i \epsilon\right)}^{-1}\nonumber \\
& & 
\cdot e^{ i \omega (t' - t)} \sin\frac{m \varphi}{2} \sin\frac{m \varphi'}{2} \,,
\label{eq:G:explicit}
\eeqn
where the infinitesimally small imaginary term $i \epsilon$ in the denominator guarantees that the Green's function~\eq{eq:G:general} is of the Feynman type, Eq.~\eq{eq:G:phi}: The integration contour passes below (above) the poles on the negative (positive) part of the real axis~\cite{ref:Milton:Book}; see Fig.~\ref{fig:contour}.

Next, we substitute the integral representation of the Green's function~\eq{eq:T00} into the expression for the energy density~\eq{eq:G:general} and then we calculate the mean energy density~\eq{eq:calE:R}:
\beqn
{\bar{\mathcal E}} = \int_{0}^{2 \pi } \frac{d \varphi}{2 \pi} \, \left\langle T^{00} (t,\varphi) \right\rangle\,.
\label{eq:total:energy}
\eeqn
There are various ways to perform this calculation.

\vskip 3mm

{\bf {Time--splitting regularization}} \cite{ref:Milton:Book,ref:Bordag:Book}. In order to regularize the energy~\eq{eq:total:energy} we use the Green's function~\eq{eq:G:general} with 
\beqn
t - t' = \delta t\,,
\label{eq:time:splitting}
\eeqn
where $\delta t$ is a small but finite parameter (to be later sent to zero). The regularized zero-point energy then reads as follows:
\beqn
{\bar{\mathcal E}} & = & \frac{\Omega_{0}}{4 \pi} \int\limits_{-\infty}^{+\infty} \frac{d \omega}{2 \pi i}  \sum_{m=1}^{\infty} \frac{\left(\frac{m^{2}}{4 R^{2}} + \omega^{2} \right) e^{- i \omega \delta t}}{\frac{m^{2}}{4 R^{2}} - \omega^{2} - i \epsilon} \nonumber \\
& = & \frac{1}{8 \pi R^{2}} \sum_{m=1}^{\infty} m\, e^{- i m \delta t/ (2 R)} \,.
\label{eq:T00:reg}
\eeqn
We have taken the time--splitting parameter $\delta t$ to be positive, $\delta t > 0$ (one could equivalently use $\delta t < 0$ as well), and closed the contour in Fig.~\ref{fig:contour} at the infinitely large semicircle in the lower half plane. Thus, only positive--valued poles at the real axis enter the expression for the energy density~\eq{eq:T00:reg}, and the corresponding residues are as follows:
\beqn
\mathop{\text{res}}_{\omega = \frac{m}{2 R}} \frac{\omega^{2} e^{- i \omega \delta t}}{\frac{m^{2}}{4 R^{2}} - \omega^{2} - i \epsilon} = - \frac{m}{4 R} e^{- i m \delta t/ (2 R)}\,.
\eeqn

The sum in Eq.~\eq{eq:T00:reg} may be taken explicitly, and it gives in the limit $\delta t \to 0$:
\beqn
{\bar{\mathcal E}} = - \frac{1}{2\pi \delta t^{2}} - \frac{1}{96 \pi R^{2}} + O\left(\delta t^{2}/R^{4}\right)\,.
\label{eq:calE}
\eeqn
The first term in this expression is a divergent quantity which is independent on the circle's radius $R$. Therefore the first term corresponds for the divergent energy density of the zero--point fluctuations. The second term in Eq.~\eq{eq:calE} gives us a finite Casimir energy density of the vacuum fluctuations due to the vacuum fluctuations, and other terms in Eq.~\eq{eq:calE} vanish in the limit of vanishing time splitting, $\delta t \to 0$.

Finally, we get the following expression for the zero--point energy density of the scalar field at the nonrotating circle with the Dirichlet cut:
\beqn
{\bar{\mathcal E}}^{\mathrm{phys}} = - \frac{1}{96 \pi R^{2}}\,.
\label{eq:T00:reg:2}
\eeqn
This energy density turns zero in the limit of the infinitely large circle, $R \to \infty$. The corresponding total vacuum energy~\eq{eq:Casimir:energy} is given by the following expression:
\beqn
{E}^{\mathrm{phys}} = 2 \pi R {\bar{\mathcal E}}^{\mathrm{phys}} = - \frac{1}{48 R} \,.
\label{eq:Casimir:total}
\eeqn
This result is known as the Casimir energy of the string of the length $l = 2 \pi R$~\cite{ref:Milton:Book,Luscher:1980fr}. 

\vskip 3mm

{\bf {Time--splitting regularization: explicit Green's function.}} It is very convenient to compute the local energy density via explicit calculation of the Green's function~\eq{eq:G:explicit}: 
\beqn
G(t,t';\varphi,\varphi') & = & \sum_{m=1}^{\infty}  \frac{i}{\pi m} 
\sin\frac{m \varphi}{2} \sin\frac{m \varphi'}{2}  e^{- i \frac{m \Omega_{0}}{2} |(t - t')|}\,.
\nonumber \\
& = & \frac{i}{\pi} {\mathcal{G}}\left(\frac{\varphi}{2},\frac{\varphi'}{2}, \frac{| (t - t')|}{2 R} \right)\,,
\label{eq:explicit:G:1}
\eeqn
where the contour of integration (Fig.~\ref{fig:contour}) is closed by a large semicircle in the upper (lower) half plane if $t'>t$ ($t'<t$) so that the contour encloses only the poles located at the negative (positive) part of the real axis.

The function ${\mathcal{G}}$ in Eq.~\eq{eq:explicit:G:1} is defined as follows:
\beqn
{\mathcal{G}}(x,y,z) & = & \sum_{m=1}^{\infty} \frac{\sin(m x) \sin(m y) }{m} e^{- i m z} 
\label{eq:G:cal} \\
& = & \frac{1}{4} \ln \frac{\left[1 {-} e^{i (x + y - z)}\right] 
\left[ 1 {-} e^{- i (x + y + z)}\right]}{\left[ 1 {-} e^{i (x - y - z)}\right] \left[ 1 {-} e^{i (- x + y - z) }\right]} \nonumber \\
& \equiv & \frac{1}{4} \ln \left|\frac{\cos(x+y) - \cos z}{\cos(x-y) - \cos z} \right|  \nonumber \\
& & - \frac{i}{8} \bigl( {[z-x-y]}_{2\pi} + {[z+x+y]}_{2\pi} \nonumber \\
& & \hskip6.5mm - {[z-x+y]}_{2\pi} - {[z+x-y]}_{2\pi} \bigr)\,, \nonumber
\eeqn
where we have used the following formula\footnote{The cut of the logarithmic function $\ln z$ is located on the real axis at ${\mathrm{Re}} \, z<1$~\cite{ref:Abramowitz}.}:
\beqn
- \sum_{m=1}^{\infty} \frac{e^{i m x}}{m} = \ln\left(1{-} e^{i x}\right) \equiv \ln \left(2 \left|\sin \frac{x}{2}\right|\right) + \frac{i}{2} \left( {[x]}_{2\pi} {-} \pi \right)\!.
\qquad \nonumber
\eeqn

By using the explicit representation of the Green's function~\eq{eq:explicit:G:1} the local energy density can be computed straightforwardly:
\beqn
& & \left\langle T^{00}(t,\varphi) \right\rangle = \frac{1}{2 i}  \lim_{t' \to t} \lim_{\varphi' \to \varphi} 
\left( \partial_{t} \partial'_{t} + \frac{1}{R^{2}} \partial_{\varphi} \partial'_{\varphi} \right) 
\label{eq:explicit:time:splitting} \\
& & \hskip 15mm G(t,t';\varphi,\varphi')  = - \frac{1}{2 \pi} \lim_{t' \to t} \frac{1}{(t' - t)^{2}} - \frac{1}{96 \pi R^{2}}\,.
\nonumber
\eeqn
We again have arrived to Eq.~\eq{eq:T00:reg:2}. Notice that the local energy density~\eq{eq:explicit:time:splitting} turns out to be independent on the angular variable~$\varphi$.

\vskip 3mm

{\bf Formal {${\boldsymbol{\zeta}}$--function regularization.}} One can also formally start from Eq.~\eq{eq:T00:reg} at $\delta t = 0$ 
and regularize the divergent sum over the modes~$m$,
\beqn
{\bar{\mathcal E}}(s) = \frac{1}{8 \pi R^{2}} \zeta(s) \,,
\label{eq:T00:reg:zeta:1}
\eeqn
via the Riemann's $\zeta$ function:
\beqn
\zeta(s) = \sum_{m=1}^{\infty} m^{-s}\,.
\label{eq:zeta:function}
\eeqn
The regularization of the sum in Eq.~\eq{eq:T00:reg:zeta:1} may be done via an analytical continuation of the $\zeta$ function,
\beqn
\sum_{m=1}^{\infty} m \overset{\text{reg}}{=} \lim_{s\to -1} \zeta(s)  = \zeta(-1) = - \frac{1}{12}\,,
\label{eq:zeta:1}
\eeqn
which -- after substitution to Eq.~\eq{eq:T00:reg:zeta:1} -- again gives the expression for the known vacuum energy density~\eq{eq:T00:reg:2}.

\subsubsection{$\zeta$--function regularization of explicit sum over the modes}
\label{sec:static:zeta}

A quick derivation of the vacuum energy can also be done via a direct $\zeta$--function regularization of the sum~\eq{eq:calE:R} over the energies of individual field's fluctuations.

First, we should find the energy spectrum of the circle with the cut. The energy eigenmodes satisfy 
the classical equation of motion of the Lagrangian~\eq{eq:circle:L},
\beqn
\left(\frac{\partial^{2}}{\partial t^{2}} - \frac{1}{R^{2}} \frac{\partial^{2}}{\partial \varphi^{2}}\right) \phi^{(0)}_{m} (t,\varphi) = 0\,,
\label{eq:EOM:0}
\eeqn
and the boundary condition~\eq{eq:cirle:D:0}. As usual, the corresponding real-valued eigenfunctions can conveniently be combined into the complex valued function,
\beqn
\phi^{(0)}_{m}(t,\varphi) & \equiv & \phi^{(0),C}_{m,\omega} (t,\varphi) - i \phi^{(0),S}_{m} (t,\varphi)\,.
\eeqn
The eigenfunctions $\phi^{(0)}_{m}$ and eigenenergies $\epsilon^{(0)}_{m}$ are labeled by a positive integer number~$m$,
\beqn
\phi^{(0)}_{m} (t,\varphi) & = & e^{- i \varepsilon^{(0)}_{m} t} \chi^{(0)}_{m} (\varphi) \,,
\label{eq:phi:static} \\
\epsilon^{(0)}_{m} & = & \frac{m}{2 R}\,, 
\label{eq:epsilon:0} \\
m & = & 1,2,3,\dots\,, \nonumber
\eeqn
where the spatial eigenfunction $\chi^{(0)}_{m}$ is given in Eq.~\eq{eq:chi:static}.

The spatial eigenmodes~\eq{eq:chi:static} are orthonormal,
\beqn
R \int_{0}^{2 \pi} d \varphi \, \chi^{(0)\dagger}_{m} (t,\varphi) \chi^{(0)}_{m'} (t,\varphi) = \delta_{m m'}\,,
\eeqn
and their basis is complete:
\beqn
\sum_{m=1}^{\infty} \chi^{(0)}_{m_{1}}(\varphi)\chi^{(0)\dagger}_{m_{2}}(\varphi') & =  & \frac{1}{R} \sum_{k \in \Z} \left[\delta(\varphi - \varphi' + 4 \pi k) \right. \nonumber \\
& & \left. -\delta(\varphi + \varphi' + 4 \pi k)\right] \,.
\label{eq:completeness:static}
\eeqn
The second $\delta$--function in the right hand side of Eq.~\eq{eq:completeness:static} guarantees that the sum in the left hand side is zero at the position of the Dirichlet cut, $\varphi = 0,2\pi$, and at $\varphi' = 0,2\pi$. This property is expected because of the Dirichlet boundary condition~\eq{eq:cirle:D:0}.

Second, we should sum all energies of the individual fluctuations~\eq{eq:epsilon:0} and get the zero--point energy of the scalar field at the circle with the Dirichlet cut: 
\beqn
E = \frac{1}{2} \sum_{m=1}^{\infty} \varepsilon^{(0)}_{m} \equiv \frac{\Omega_{0}}{4} \sum_{m=1}^{\infty} m\,.
\label{eq:E:static:naive}
\eeqn
Since this sum is a divergent quantity, we following the standard approach~\cite{ref:Milton:Book,ref:Bordag:Book,ref:Elizalde,ref:Kirsten,Elizalde:1988rh}, 
and we regularize the sum~\eq{eq:E:static:naive} by using the $\zeta$ regularization:
\beqn
E(s) = \frac{1}{2} \sum_{m=1}^{\infty} \bigr[\omega^{(0)}_{m} \bigl]^{-s} = \frac{2^{s-1}}{\Omega_{0}^{s}} \sum_{s=1}^{\infty} m^{-s} \equiv \frac{2^{s-1}}{\Omega_{0}^{s}} \zeta(s)\,, \qquad
\eeqn
where $\zeta$ is the Riemann's $\zeta$ function~\eq{eq:zeta:function}. 

In order to calculate the physical part of the regularized energy~\eq{eq:E:static:naive} we use, as usual, the analytical continuation of the $\zeta$ function to the point $s=-1$, which gives us $\zeta(-1) = -1/12$; see Eq.~\eq{eq:zeta:1}. Then the physical part of the energy density~\eq{eq:E:static:naive} becomes a finite quantity which coincides with the result obtained in all other approaches~\eq{eq:Casimir:total}.

\subsection{Rotating circle with the Dirichlet cut}
\label{sec:rotating:D}

In the previous section, we have overviewed various well-known approaches to standard calculation of the zero--point energy of a scalar field in the circle with the Dirichlet cut, and we have highlighted the equivalence of these approaches. Although there are various ways to calculate the zero--point energy for the case of the uniformly rotating circle, below we chose the simplest and most straightforward approach based on the explicit calculation of the Green's function and the time--splitting regularization. We recover the result of Ref.~\cite{ref:I} where a zeta--function method~\cite{ref:Elizalde,ref:Kirsten,Elizalde:1988rh} was used.

\subsubsection{Eigenfunctions in the rotating circle}

Let us consider the circle with the Dirichlet cut which rotates uniformly about its central point with a uniform angular velocity $\Omega$; see Fig.~\ref{fig:setup}. The rotation leads to the following time--dependent boundary condition:
\beqn
\phi (t,\varphi)\bl_{\varphi = {[\Omega t]}_{2\pi}}  = 0\,,
\label{eq:cirle:D:cond}
\eeqn
where
\beqn
{[x]}_{2 \pi} = x + 2 \pi n\,, \qquad n \in \Z\,, \qquad 0 \leqslant {[x]}_{2 \pi} < 2 \pi, \qquad
\label{eq:modulus}
\eeqn
denotes the modulo operation with the base of $2 \pi$. We also impose the physical condition that the velocity of the Dirichlet cut in the laboratory frame should not exceed the speed of light, $|\Omega R| < 1$. In the nonrotating limit, $\Omega = 0$, Eq.~\eq{eq:cirle:D:cond} reduces to Eq.~\eq{eq:cirle:D:0}.

As in the nonrotating case, the energy density is given by Eq.~\eq{eq:T00}, where the Green's function~\eq{eq:G:general} is expressed via the solutions of Eq.~\eq{eq:phi:solutions} but now the time--dependent boundary conditions~\eq{eq:cirle:D:cond} should be used. In the laboratory frame the corresponding wave functions are
\beqn
\phi_{m,\omega}(t,\varphi) & = & \sqrt{\frac{1}{\pi R}} \sin \Bigl[\frac{m}{2} {[\varphi - t \Omega]}_{2 \pi} \Bigr] 
\nonumber \\ & & \cdot 
\exp\biggl\{ - i \omega \biggl(t - \frac{\Omega R^{2}\, {[\varphi - t \Omega]}_{2 \pi}}{1 - \Omega^2 R^{2}} \biggr)\biggr\}, \quad
\label{eq:phi:m}
\eeqn
with $m = 1,2,3, \dots$. One can show that these wave functions satisfy both the orthonormality~\eq{eq:orthonormal:static} and completeness~\eq{eq:complete:static} conditions.

The corresponding eigenvalues are as follows:
\beqn
\lambda_{\omega,m} = \frac{1 - \Omega^{2} R^{2}}{4 R^{2}} m^{2} - \frac{\omega^{2}}{1 - \Omega^{2} R^{2}} \,. \qquad 
\label{eq:lambda:omega:m}
\eeqn

Technically, a simple derivation of the wave functions~\eq{eq:phi:m} can be done, for example, by changing the coordinates from the laboratory frame, $(t,\varphi)$, to the rotating frame, $(\vartheta,\theta)$, in which our object is static: 
\beqn
\vartheta = t\,, \qquad \theta = \varphi - \Omega t\,.
\label{eq:coordinates}
\eeqn

In the rotating frame the Dirichlet boundary condition~\eq{eq:cirle:D:cond} takes a simpler form:
\beqn
{\widetilde{\phi}} (\vartheta,\theta)\bl_{\theta = 0} = 0\,,
\label{eq:cirle:D:cond2}
\eeqn
where ${\widetilde{\phi}}(\vartheta,\theta) \equiv \phi(\vartheta,\theta + \Omega t)$. The eigenvalue equation~\eq{eq:phi:solutions} takes a new form:
\beqn
& & \left[\left(\frac{\partial}{\partial \vartheta} - \Omega \frac{\partial}{\partial \theta}  \right)^{2} - \frac{1}{R^{2}} \frac{\partial^{2}}{\partial \theta^{2}}\right]  
{\widetilde{\phi}}_{\omega,m}(\vartheta,\theta) \nonumber \\
& & \hskip 43mm = \lambda_{\omega,m} \, {\widetilde{\phi}}_{\omega,m}(\vartheta,\theta)\,, \qquad
\label{eq:S:dA}
\eeqn
Solving Eqs.~\eq{eq:cirle:D:cond2} and \eq{eq:S:dA} and coming back to the laboratory frame one gets the wave functions~\eq{eq:phi:m}.

\vskip 3mm
~
\subsubsection{Time--splitting: Explicit Green's function calculation}
\label{sec:rotating:Green:time:splitting}

We substitute the wave function~\eq{eq:phi:m} into the general expression~\eq{eq:G:general} and repeat all the steps which led us from Eq.~\eq{eq:G:general} to the derivation of the Green's function in the nonrotating case~\eq{eq:explicit:G:1}. To this end we notice that the wave function of the rotating circle~\eq{eq:phi:m} is very similar (up to a redefinition of the time and angular coordinates) to the wave function \eq{eq:phi:omega} and \eq{eq:chi:static} of the nonrotating circle. The same is true (up to a recalling of $\omega$ and of a prefactor in the Green's function) for the eigenvalues\eq{eq:lambda:omega:m} and \eq{eq:eigenvalues}. As a result, we get the following Green's function for the scalar field on a circle with the Dirichlet cut:
\begin{widetext}
\beqn
G_{\Omega}(t,t';\varphi,\varphi') = \frac{i}{\pi} {\mathcal{G}}\left(\frac{{[\varphi - \Omega t]}_{2\pi}}{2},\frac{{[\varphi' - \Omega t']}_{2\pi}}{2},
\frac{\left|(1 - \Omega^{2} R^{2}) (t - t') - \Omega R^{2} \left({[\varphi - \Omega t]}_{2\pi} - {[\varphi' - \Omega t']}_{2\pi}\right) \right|}{2 R} \right)\,,
\label{eq:G:Omega}
\eeqn
\end{widetext}
where the function ${\mathcal G}(x,y,z)$ is given in Eq.~\eq{eq:G:cal}.

Then the local zero--point energy density is:
\beqn
\left\langle T^{00}(t,\varphi) \right\rangle & = & \frac{1}{2 i}  \lim_{t' \to t} \lim_{\varphi' \to \varphi} 
\label{eq:T00:gen} \\
& & \left( \partial_{t} \partial'_{t} + \frac{1}{R^{2}} \partial_{\varphi} \partial'_{\varphi} \right) G(t,t';\varphi,\varphi') \nonumber \\
& =& - \frac{1}{2 \pi } \lim_{t' \to t} \frac{1}{(t' - t)^{2}} - \frac{1 + \Omega^{2} R^{2}}{96 \pi R^{2}}
\,.
\nonumber
\eeqn
The divergent part of this expression is equivalent to the one in the static case~\eq{eq:explicit:time:splitting}. This divergence depends neither on the radius of the circle nor on the angular frequency of rotation and therefore it does not contribute to the physical part of the rotational energy density of zero--point (ZP) fluctuations:
\beqn
{\mathcal E}^{\mathrm{ZP}}_{\Omega} (t,\varphi) \equiv {\left\langle T^{00}(t,\varphi) \right\rangle}^{\mathrm{phys}}  = - \frac{1 + \Omega^{2} R^{2}}{96 \pi R^{2}}\,.
\label{eq:T00:rotating}
\eeqn
The physical energy density of the rotating device depends neither on time $t$ nor on the angular variable $\varphi$ so that for the rotating device with the Dirichlet cut the energy density of the zero--point fluctuations~\eq{eq:T00:rotating} is the explicitly time--independent quantity:
\beqn
\frac{\partial}{\partial t} {\mathcal E}^{\mathrm{ZP}}_{\Omega} (t,\varphi) \equiv 0\,.
\eeqn
For a static device, $\Omega = 0$, Eq.~\eq{eq:T00:rotating} reduces to the known result~\eq{eq:T00:reg:2}.

Finally, we get the following exact relativistic expression for the  total zero--point energy of the rotating circle~\cite{ref:I}:
\beqn
E^{\mathrm{ZP}}_{\Omega} & \equiv & R \int\limits_{0}^{2 \pi} d \varphi \, {\mathcal E}^{\mathrm{ZP}}_{\Omega} (t,\varphi) = - \frac{1 + R^{2} \Omega^{2}}{48 R}. \qquad 
\label{eq:ZP:B0}
\eeqn
We would like to stress that Eq.~\eq{eq:ZP:B0} provides us with the {\it exact} relativistic expression for an infinitely thin circle with an infinitely thin Dirichlet cut which rotates with the angular frequency $\Omega$. Equations \eq{eq:T00:rotating} and \eq{eq:ZP:B0} are valid provided the Dirichlet cut moves with a subluminal velocity, $|\Omega R| < 1$.

\subsubsection{Physical features of the device}

There are three important general physical features of our device. 

{\bf First}, the rotational energy is a negative-values quantity, the absolute value of which increases quadratically with the angular frequency $\Omega$. One can define the ``moment of inertia'' of the zero--point fluctuations\footnote{We put the term moment of inertia in quotation marks because we have identified it as if the system is nonrelativistic. Indeed, the rotational part of the energy in Eq.~\eq{eq:ZP:B0} resembles a nonrelativistic behavior while it is computed for the zero--point fluctuations corresponding to relativistic massless particles. The system is neither fully relativistic nor fully nonrelativistic. In this unusual case, Eq.~\eq{eq:ZP:B0} should be considered as a definition of the moment of inertia of the zero--point fluctuations.}:
\beqn
I^{\mathrm{ZP}} \equiv \frac{\partial^{2}}{\partial \Omega^{2}} E^{\mathrm{ZP}}_{\Omega} = - \frac{\hbar R}{24 c}\,,
\label{eq:I:ZP:no:enhancement}
\eeqn
where we have restored the Planck constant $\hbar$ and the speed of light $c$. 

The negative value of the moment of inertia of the zero--point fluctuations is a natural fact. Indeed, the inertial mass corresponding to the Casimir energy $E_{c}$ is always $E_{c}/c^{2}$ regardless of the sign of the Casimir energy itself~\cite{Fulling:2007xa}, so that the negative mass should have a negative moment of inertia. If the device itself were massless then the ground state of the device would correspond to the permanently rotating state due to the negative moment of inertia of the zero--point fluctuations. 

In our simplest example, the zero--point moment of inertia~\eq{eq:I:ZP:no:enhancement} is tiny:
\beqn
I^{\mathrm{ZP}} = - 1.5 \times 10^{- 44} \cdot \left( \frac{R}{{\mathrm{m}}} \right) \cdot \mathrm{kg} \, \mathrm{m}^{2}\,.
\label{eq:I:ZP:phys}
\eeqn
It is too small~\eq{eq:I:ZP:phys} to overcome a classical moment of inertia of a real device. However, in the next Section we show that the negative moment of inertia can be drastically enhanced by an external magnetic field if the massless particles are electrically charged, so that a fabrication of a permanently rotating device may become closer to reality.

{\bf Second}, the zero--point rotational energy~\eq{eq:ZP:B0} is unbounded from below at the relativistically large angular frequencies $|\Omega| \to 1/R$. This feature is an artifact which emerges due to our mathematical simplification which assumes that the thickness of our circle is infinitely small. One can show that for spatially extended systems -- such as a cylinder -- the rotational zero--point energy has its minimum at finite values of the angular frequency. In this case the dependence of the zero--point energy on the rotational frequency has the form of a double-well potential~\cite{ref:I} with nontrivial minima at $\Omega \neq 0$.

{\bf Third}, it is important to stress that no transition from a nonrotating state ($\Omega = 0$) to a rotating one ($\Omega \neq 0$) may occur for an {\it isolated} device because the angular momentum is a conserved quantity. A transition towards the rotating ground state may be realized only in the presence of an environment. 

In order to support this statement we calculate the force ${\mathcal F}^{\mathrm{ZP}}$ which is experienced by the Dirichlet cut due to the zero--point fluctuations. Following the line of arguments of Ref.~\cite{ref:Milton:2004}, the force can be expressed as follows:
\beqn
{\mathcal F}^{\mathrm{ZP}} & = & \frac{1}{R^{2}} \biggl( \left\langle T^{\varphi\varphi}(t,\varphi) \right\rangle \bl_{\varphi = {[\Omega t -0 ]}_{2\pi}} \nonumber \\
& & \hskip 4.5mm - \left\langle T^{\varphi\varphi}(t,\varphi) \right\rangle \bl_{\varphi = {[\Omega t + 0 ]}_{2\pi}} \biggr)\,,
\eeqn
where the difference of the expectation value of the component $T^{\varphi\varphi}(t,\varphi)$ is taken at the left and at the right sides of the cut~\eq{eq:cirle:D:cond}. 

According to the general expression~\eq{eq:Tmunu}, 
\beqn
\left\langle T^{\varphi\varphi}(t,\varphi) \right\rangle \equiv - R^2 \left\langle T^{00}(t,\varphi) \right\rangle\,.
\eeqn
This relation is a natural fact because the theory~\eq{eq:circle:L} is conformal at both classical and quantum levels (there is no conformal anomaly), so that the expectation value of the trace of the stress--energy tensor should be zero. Taking into account the fact that the expectation value of the $T^{00}$ component~\eq{eq:T00} does not depend on angular variable~\eq{eq:explicit:time:splitting}, we obtain that the zero--point fluctuations produce no force on the Dirichlet cut:
\beqn
{\mathcal F}^{\mathrm{ZP}} = 0\,.
\eeqn

Thus, the isolated device is not self--accelerating because of the conservation of the angular momentum. The absence of the force on the Dirichlet cut is one of the major differences of the rotational vacuum effect and the conventional Casimir effect: Despite the fact that the ground state of the device corresponds to a rotating state, the device -- even if it is not residing in its ground state -- will not self--accelerate unless it exchanges the angular momentum with an environment or, equivalently, unless it emits the extra angular momentum via radiation of, e.g., a photon.

\section{Conducting circle with \newline the Dirichlet cut in a magnetic field}
\label{sec:charged:1d} 

\subsection{The device}

\begin{figure}[!htb]
\begin{center}
\includegraphics[scale=0.2,clip=false]{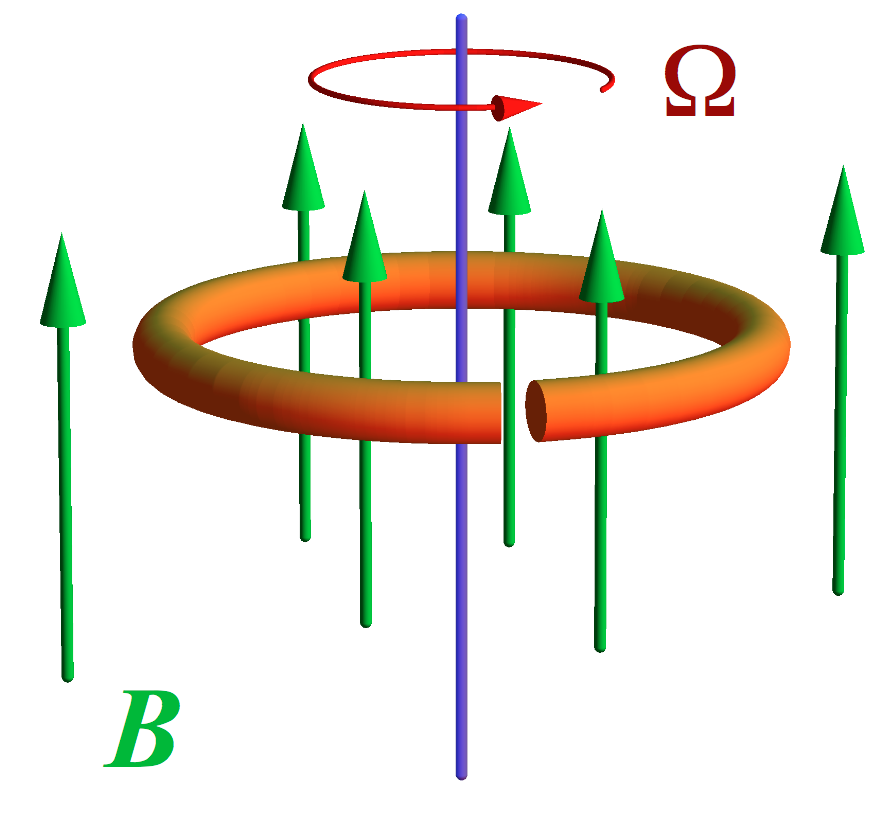}
\end{center}
\vskip -5mm
\caption{A simplest device which demonstrates an enhancement of the negative moment of inertia of zero--point fluctuations due to the magnetic field $B$. The circle supports electrically charged and massless scalar excitations while the cut imposes a Dirichlet boundary condition on these excitations.}
\label{fig:circle}
\end{figure}

Consider an electrically charged massless scalar field $\Phi = \Phi(t,\varphi)$ which is defined, as in the previous section, on a circle with a fixed radius $R$; see Fig.~\ref{fig:circle}. The field $\Phi$ is electrically charged and it is interacting with a background electromagnetic field $A_{\mu}$. The corresponding Lagrangian is as follows:
\beqn
\cL & = & \left[D_{\mu} \Phi\right]^{*} D^{\mu} \Phi \nonumber \\
& \equiv & \left[D_{t} \Phi\right]^{*} D_{t} \Phi - \frac{1}{R^{2}} \left[D_{\varphi} \Phi\right]^{*} D_{\varphi} \Phi\,,
\label{eq:circle:L:B} \quad
\eeqn
where $D_{\mu} = \partial_{\mu} - i e A_{\mu}$ is the covariant derivative.

As usual, we consider the simplest, Dirichlet boundary condition at the position of the cut. The circle rotates uniformly about its central point with an angular velocity $\Omega$, so that the rotation leads to the following time--dependent boundary condition at the position of the cut:
\beqn
\Phi (t,\varphi)\bl_{\varphi = {[\Omega t]}_{2\pi}}  = 0\,,
\label{eq:cirle:D:cond:B}
\eeqn
where the modulo operation ${[\dots]}_{2 \pi}$ is defined in Eq.~\eq{eq:modulus}.

We consider our circle in a background of a uniform (i.e., space- and time-independent) magnetic field $B$. Since the model~\eq{eq:circle:L:B} is invariant under Maxwellian $U(1)$ gauge transformations,
\beqn
U(1): \quad \Phi \to e^{i e \omega} \Phi\,, \quad A_{\mu} \to A_{\mu} + \partial_{\mu} \omega\,,
\eeqn
it is convenient to choose a gauge where the gauge field has the following form\footnote{We work with the cylindrical coordinates,
$\boldsymbol{A} = A_{\rho}\hat{\boldsymbol{\rho}} + A_{\varphi} \hat{\boldsymbol{\varphi}} + A_{z} \hat{\boldsymbol{z}}$, where $\hat{\boldsymbol{\rho}}$, $\hat{\boldsymbol{\varphi}}$, and $\hat{\boldsymbol{z}}$ are unit orthogonal vectors. We consider the vacuum of the scalar particles at the circle and not in the interior or exterior of the circle, so that $A_{\varphi}$ and $A_z$ components are completely irrelevant for our problem, while the behavior of the $A_{\varphi} = A_{\varphi}(\rho)$ component is relevant only at $\rho = R$.}:
\beqn
A_{\varphi} = \frac{\gamma_{B}}{e} \,, \qquad A_{t} = 0 \,, \qquad A_{\rho} = 0\,, \qquad A_{z} = 0\,, \qquad
\label{eq:A}
\eeqn
where 
\beqn
\gamma_{B} = \frac{e F_{B}}{2 \pi}\,,
\label{eq:gamma}
\eeqn
is a constant and $F_{B}$ is the flux of the magnetic field $B$ which pierces the surface $S$ spanned on the circle $C \equiv  \partial S$:
\beqn
F_{B} = \oiint_{S} d^{2} {\boldsymbol{s}} \cdot {\boldsymbol{B}} \equiv \oint_{C} d {\boldsymbol{x}} \cdot {\boldsymbol{A}} = R \int\limits_{0}^{2\pi} d \varphi \, A_{\varphi}\,.
\label{eq:F}
\eeqn

Below, we calculate the energy of zero--point fluctuations for this device following the line of previous sections.

\subsection{Energy density of zero-point fluctuations}

\subsubsection{The eigensystem}

The eigensystem problem for the Lagrangian~\eq{eq:circle:L:B} with the gauge field~\eq{eq:A} is the following:
\beqn
& & \left[\frac{\partial^{2}}{\partial t^{2}} {-} \frac{1}{R^{2}} \left(\frac{\partial}{\partial \varphi}  - i \gamma_{B} \right)^{2}\right] \Phi_{\omega,m}(t,\varphi) \nonumber \\
& & \hskip 30mm =  \Lambda_{\omega,m} \, \Phi_{\omega,m}(t,\varphi)\,. \quad
\label{eq:phi:solutions:B}
\eeqn

The eigenvalues and eigenfunctions of Eq.~\eq{eq:phi:solutions:B} with the boundary condition~\eq{eq:cirle:D:cond:B} are, respectively, as follows:
\beqn
\Lambda_{\omega,m} & = & \frac{1 - \Omega^{2} R^{2}}{4 R^{2}} m^{2} - \frac{(\omega + \gamma_{B} \Omega)^{2}}{1 - \Omega^{2}  R^{2}} \,. \qquad
\label{eq:Lambda:omega:m} \\
\Phi_{\omega,m}(t,\varphi) & = & \sqrt{\frac{1}{\pi R}} \sin \Bigl(\frac{m}{2} {[\varphi - t \Omega]}_{2 \pi} \Bigr) \label{eq:phi:m:B} \\
& & \hskip -7mm \cdot \exp\biggl\{ - i \omega t + i \frac{\gamma_{B} + \omega\, \Omega R^{2}}{1 - \Omega^2 R^{2}} {[\varphi - t \Omega]}_{2 \pi} \biggr\},
\nonumber
\eeqn
where $m = 1,2,3, \dots$. The wave functions are orthonormal and they form a complete basis.

\subsubsection{The energy density}
The local energy density of the zero--point fluctuations ${\mathcal E}(x)$ is given by the vacuum expectation value~\eq{eq:E:T00} of the $T^{00}$ component of the stress--energy tensor.
This expectation value can be computed by using a Feynman--type Green function 
\beqn
G(x,x') = i \left\langle {\mathrm T} \Phi (x) \Phi^{*} (x') \right\rangle\,,
\label{eq:G:phi:B}
\eeqn
via the following familiar relation:
\beqn
\left\langle T^{\mu\nu} (x) \right\rangle & = & \left( D^{\mu} D'^{\nu *} + D^{\nu} D'^{\mu *} \right. \nonumber \\
& & - \left.g^{\mu\nu} D^{\lambda} D^{\prime *}_{\lambda}\right) \frac{1}{i} G(x,x') \bl_{x \to x'}\,,
\label{eq:Tmunu:B}
\eeqn
so that
\beqn
\left\langle T^{00} (t,\varphi) \right\rangle & = & \left[\frac{\partial}{\partial t} \frac{\partial}{\partial t'} + \frac{1}{R^{2}} \left(\frac{\partial}{\partial \varphi} - i \gamma_{B}\right) 
\left(\frac{\partial}{\partial \varphi'} + i \gamma_{B} \right) \right] \nonumber \\
& &  
\frac{1}{i} G(t,t';\varphi,\varphi') \bl_{{}^{t \to t'}_{\varphi \to \varphi'}}\,. \qquad
\label{eq:T00:B}
\eeqn

The Green's function can be expressed via the eigenfunctions~\eq{eq:phi:m:B} and eigenvalues~\eq{eq:Lambda:omega:m} similarly to Eq.~\eq{eq:G:phi}:
\beqn
G_{\Omega,B}(t,t';\varphi,\varphi') & = & \int\limits_{-\infty}^{+\infty} \frac{d \omega}{2 \pi}  \sum_{m=1}^{\infty}
\frac{\Phi_{\omega,m}(t,\varphi) \Phi^*_{\omega,m}(t',\varphi')}{\Lambda_{\omega,m} - i \epsilon} \,, \qquad
\label{eq:G:general:B}
\eeqn

The positions of poles $\omega = \omega_{m}$ are determined by the following equation: $\Lambda_{\omega,m} = 0$. According to Eq.~\eq{eq:Lambda:omega:m} the poles are located at the real axis:
\beqn
\omega_{m} = \frac{1 - \Omega^{2} R^{2}}{2 R} m - \gamma_{B} \Omega \equiv \mu_{m} - \gamma_{B} \Omega\,,
\label{eq:omega:m:pm}
\eeqn
where $m \in \Z$ is an integer number.

\begin{figure}[!thb]
\begin{center}
\includegraphics[scale=0.25,clip=false]{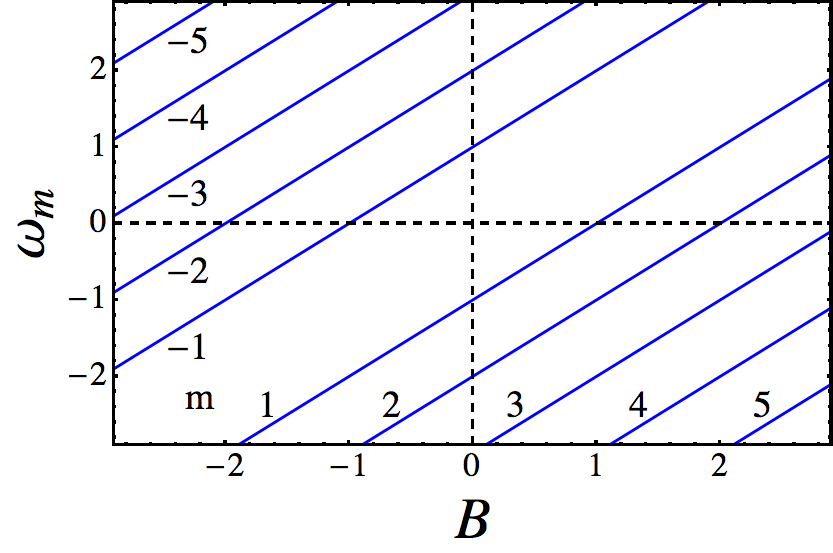}
\end{center}
\vskip -5mm
\caption{Schematic illustration of the positions of the poles~\eq{eq:omega:m:pm} in the Green's function~\eq{eq:G:general:B} vs the magnetic field~$B$ (arbitrary units) for a clockwise rotation ($\Omega < 0$). For a coun\-ter\-clock\-wise rotation ($\Omega < 0$), the slopes are negative.}
\label{fig:levels}
\end{figure}

An important novelty of Eq.~\eq{eq:omega:m:pm} is that the positions of the poles are no longer symmetric with respect to the reflections $\omega \to - \omega$ due to the presence of the flux of the magnetic field~\eq{eq:gamma}:
\beqn
F_{B} = \frac{2 \pi \gamma_{B}}{e}\,.
\eeqn
As the magnetic flux increases at nonzero angular frequency $\Omega$, some of the poles~\eq{eq:omega:m:pm} may cross the origin, $\omega = 0$, coming from a negative part of the real axis to the positive part and vice versa, as illustrated in Fig.~\ref{fig:levels}.

The number of poles which have crossed (due to the presence of the magnetic field $B$) the origin in the negative direction is
\beqn
M_{\Omega,B} & = & \left\lfloor \frac{2 \gamma_{B} \Omega R}{1 - \Omega^{2} R^{2}} \right\rfloor = \left\lfloor \frac{\Omega R}{1 - \Omega^{2} R^{2}} \frac{e F_{B}}{\pi} \right\rfloor
\label{eq:M:ceiling} 
\label{eq:M:ceiling:3} \\
 & \equiv & \left\lfloor \frac{e B \Omega R^{3}}{c^{2} - \Omega^{2} R^{2}} \, \frac{c}{\hbar} \right\rfloor \,. \nonumber
\eeqn 
Here the floor operator $\lfloor x \rfloor$ defines the largest integer which is smaller than the real number $x$ (with e.g., $\lfloor 0.1 \rfloor = 0$, $\lfloor 1.9 \rfloor = 1$, etc). The number $M_{\Omega,B}$ can be both positive and negative and it depends both on the angular frequency $\Omega$ of the circle and on the net magnetic flux $F_{B}$ which pierces the circle. The last line of Eq.~\eq{eq:M:ceiling:3} is written for a uniform magnetic field $B$, so that the magnetic flux going through the circle is
\beqn
F_{B} = \pi B R^{2} \,.
\label{eq:FB}
\eeqn
We have restored the dependence on $\hbar$ and $c$ in the last line of Eq.~\eq{eq:M:ceiling:3}.

In order to evaluate the Green's function~\eq{eq:G:general:B}, we use the following relations, valid for an arbitrary parameter $\alpha$ and an even function~$f_{m}$ 
(with $f_{m} = f_{-m}$ and $f_{0} = 0$):
\beqn
& & \int\limits_{-\infty}^{\infty} \frac{d \omega}{2\pi} \sum_{m=1}^{\infty} \frac{e^{- i \alpha \omega} f_{m}}{\Lambda_{\omega,m} - i \epsilon} 
= \int\limits_{-\infty}^{\infty} \frac{d \omega}{4\pi} \sideset{}{'}{\sum}_{m \in \Z} \frac{e^{- i \alpha \omega} f_{m}}{\Lambda_{\omega,m} - i \epsilon} \nonumber \\
& & \hskip 7mm = \frac{i e^{i \alpha \Omega \gamma_{B}}}{2 \Omega_{0}} 
\sideset{}{'}{\sum}_{m \in \Z} f_{m} \frac{e^{- i \alpha \mu_{m}}}{m} \sign(\alpha)\, \Theta(\alpha \omega_{m}) \qquad \nonumber \\
& & \hskip 7mm = \frac{i e^{i \alpha \Omega \gamma_{B}}}{2 \Omega_{0}} 
\sideset{}{'}{\sum}_{m = N_{\Omega,B}(\alpha)}^{\infty} f_{m} \frac{e^{- i |\alpha| \mu_{m}}}{m}, \qquad
\label{eq:fm:sum} 
\eeqn
where a prime in the sum indicates that the term with $m=0$ is omitted. In Eq.~\eq{eq:fm:sum} $\Theta(x)$ is the Heaviside function, and we have also defined the following integer number:
\beqn
N_{\Omega,B}(\alpha) & = & \frac{1}{2} + \left[M_{\Omega,B} + \frac{1}{2}\right] \sign(\alpha) \nonumber\\
& \equiv &  
\left\{
\begin{array}{ll}
M_{\Omega,B} +1 \,,      & \quad\alpha >0\,, \\
- M_{\Omega,B}\,, & \quad \alpha <0\,,
\end{array}
\right.
\eeqn
is an integer number.

Then we notice that for an arbitrary integer number $N$ and an arbitrary function $K_{m}$ the following relation holds:
\beqn
\sideset{}{'}{\sum}_{m = N}^{\infty} K_{m} = \sum\limits_{m=1}^{\infty} K_{m} + S\left[K_{m},N\right]\,,
\eeqn
where we have defined the following finite sum
\beqn
S\left[K_{m},N\right] = 
\left\{
\begin{array}{rl}
- \sum\limits_{m=1}^{N-1} K_{m}, &  \quad N > 1\,, \\
0, &  \quad N =0,1\,, \\
\sum\limits_{m=N}^{-1} K_{m}, &   \quad N < 0\,.
\end{array}
\right. \quad
\eeqn

Applying Eq.~\eq{eq:fm:sum} to Eqs.~\eq{eq:G:general:B} and \eq{eq:phi:m:B}, we get the following explicit representation of the Green's function:
\begin{widetext}
\beqn
G_{\Omega,B}(t,t';\varphi,\varphi') & = & \frac{i}{\pi} e^{i \left(\Omega (t-t') + {[\varphi - t \Omega]}_{2 \pi} - {[\varphi' - t' \Omega]}_{2 \pi}\right) \gamma_{B} }
\sideset{}{'}{\sum}_{m=N_{\Omega,B}(\alpha)}^{\infty}  h_{m} (t,t';\varphi,\varphi') 
\label{eq:G:Omega:B:calc} \\ & = & 
e^{i \left(\Omega (t-t') + {[\varphi - t \Omega]}_{2 \pi} - {[\varphi' - t' \Omega]}_{2 \pi}\right) \gamma_{B} } \biggl\{
G_{\Omega}(t,t';\varphi,\varphi') + \frac{i}{\pi} S \left[h_{m} (t,t';\varphi,\varphi'),
N_{\Omega,B}\bigl(\alpha (t,t';\varphi,\varphi')\bigr)\right]
 \biggr\}\,,
\nonumber
\eeqn
\end{widetext}
where $G_{\Omega}$ is the Green's function~\eq{eq:G:Omega} for the rotating circle with the Dirichlet cut in the absence of the background of the magnetic flux $F_{B}$.
In Eq.~\eq{eq:G:Omega:B:calc}, we have defined the following functions:
\beqn
\alpha (t,t';\varphi,\varphi') & = & t - t' - \frac{\Omega R^{2} \left({[\varphi - t \Omega]}_{2 \pi} - {[\varphi' - t' \Omega]}_{2 \pi}\right)}{1 - \Omega^{2} R^{2}}\,, \nonumber\\
h_{m} (t,t';\varphi,\varphi') & = & e^{- \frac{i (1 - \Omega^{2} R^{2})}{2 R} \, \left| \alpha (t,t';\varphi,\varphi') \right| m}\nonumber \\
& & \hskip -15mm \cdot \frac{1}{m} \sin \Bigl(\frac{m}{2} {[\varphi - t \Omega]}_{2 \pi} \Bigr) \sin \Bigl(\frac{m}{2} {[\varphi' - t' \Omega]}_{2 \pi} \Bigr)
\,.
\eeqn

Finally, we calculate the energy density using Eq.~\eq{eq:T00:B}:
\beqn
{\mathcal{E}}^{\mathrm{ZP}}_{\Omega,B} & \equiv & \left\langle T^{00} \right\rangle^{\mathrm{phys}}_{\Omega,B} \nonumber \\
& = & - \Bigl[1 + 6 M_{\Omega,B} (M_{\Omega,B} + 1)\Bigr]  \frac{1 + \Omega^{2} R^{2} }{48 \pi R^{2}} \,. \qquad
\label{eq:T00:rotating:B}
\eeqn
The strength of the background magnetic field $B$ enters this expression via the integer number $M_{\Omega,B}$, Eq.~\eq {eq:M:ceiling}, which depends on the angular frequency $\Omega$ as well. The energy density~\eq{eq:T00:rotating:B} does not depend on the angular coordinate $\varphi$, so that the total energy of the zero--point fluctuations~\eq{eq:Casimir:energy} is:
\beqn
E^{\mathrm{ZP}}_{\Omega,B} = - \Bigl[1 + 6  M_{\Omega,B} (M_{\Omega,B} + 1)\Bigr]  \frac{1 + \Omega^{2} R^{2}}{24 R} \,. \qquad
\label{eq:E:rotating:B}
\eeqn
Notice that in the absence of the magnetic field, $B{=}0$, Eq.~\eq{eq:E:rotating:B} equals to Eq.~\eq{eq:ZP:B0} multiplied by a factor of two because a charged field contains two degrees of freedom compared to one degree of freedom of a neutral field.

\subsection{Magnetic-field-enhanced zero-point energy}

It is clearly seen that the presence of the magnetic field enhances the negative energy of the zero--point fluctuations~\eq{eq:E:rotating:B} because the integer number $M_{\Omega,B}$ is a rising stepwise function of the magnetic field [Eq.~\eq {eq:M:ceiling}]. In order to characterize the quantity $M_{\Omega,B}$ it is convenient to introduce a characteristic frequency $\Omega_{\ch}$ of the vacuum fluctuations:
\beqn
\Omega_{\ch}(B) = \frac{\pi}{e F_{B} R} \equiv \frac{\hbar c}{e B R^{3}}\,,
\label{eq:Omega:ch:hbar} \label{eq:Omega:ch}
\eeqn
where the magnetic flux is given in Eq.~\eq{eq:FB}. Then the integer number $M_{\Omega,B}$ can be rewritten as follows:
\beqn
M_{\Omega,B} = \left\lfloor \frac{\Omega}{\Omega_{\ch}(B)} \frac{1}{1 - \Omega^{2} R^{2}}\right\rfloor \,.
\label{eq:M:ceiling:2}
\eeqn 
Notice that the characteristic frequency $\Omega_{\ch}$ is a positive number which is not limited from above.

At small angular frequencies $\Omega \sim \Omega_{\ch}$ (or, equivalently, at weak magnetic fields), the quantity $M_{\Omega,B}$ is of the order of unity and the mentioned enhancement factor for the zero--point energy~\eq{eq:E:rotating:B} is always of the order of 10. However, as the angular frequency and/or magnetic flux through the circle increase, the enhancement factor rises drastically as we will see below.

\begin{figure}[!thb]
\begin{center}
\includegraphics[scale=0.22,clip=false]{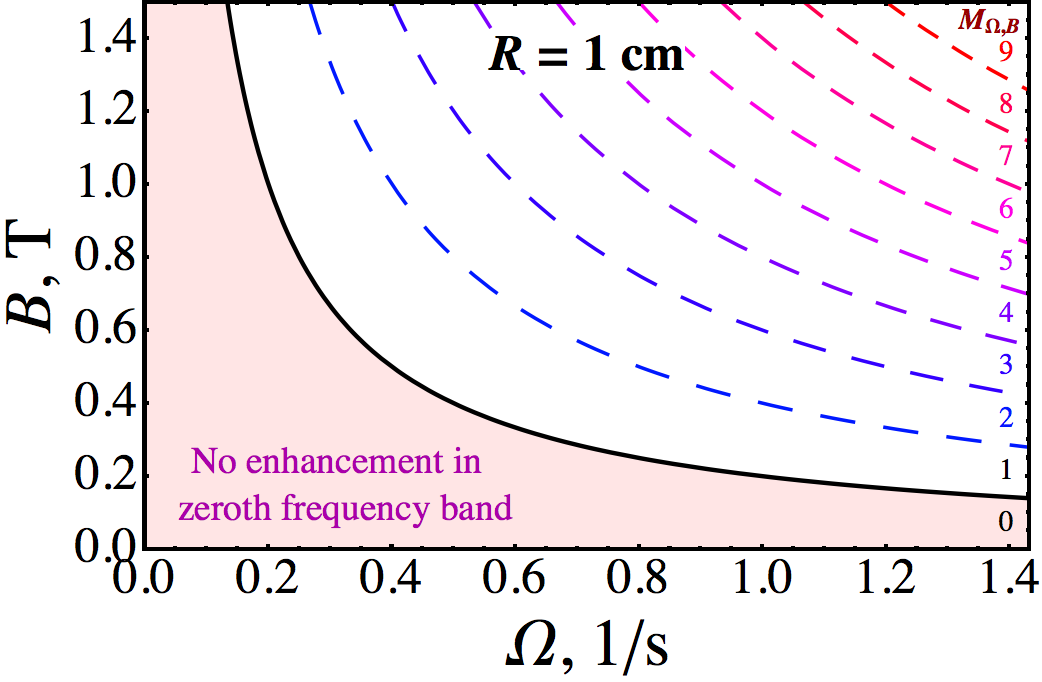}
\end{center}
\vskip -5mm
\caption{First ten ``enhancement'' bands for a circle of the radius $R=1$~cm in the plane ``angular frequency -- magnetic field''. The boundaries of the bands are defined by the relation $M_{\Omega,B} = n$ with $n \in \Z$ and $M_{\Omega,B}$ is given in Eq.~\eq{eq:M:ceiling:3} or \eq{eq:M:ceiling:phys}. }
\label{fig:bands}
\end{figure}

\begin{table*}
\begin{tabular}{r|c|c|c|c|c|l}
$B$, T & $R$ & $\Omega_{\ch}$, s${}^{-1}$ & $\tau_\ch$, s & $\Omega_{\ch} R/c$ & $f$ & \qquad Note \\
\hline
100    & 1 cm & $2 \times 10^{-3}$ & 3200 (53 min) &   $7 \times 10^{-14}$ & $2 \times 10^{27}$ & Order of max $B$ in a lab  \\
1        & 1 cm &  0.2 &  32 &  $7 \times 10^{-12}$  & $2 \times 10^{23}$ & Generic-scale $B$ and $R$ \\
1       & 1 mm & 200 & 0.03 & $7 \times 10^{-10}$  & $2 \times 10^{19}$ & 
\\
1      & 0.1 mm & $2 \times 10^{6}$ & $3 \times 10^{-5}$ & $7 \times 10^{-8}$  & $2 \times 10^{15}$ & 
Width of a human hair 
\end{tabular}
\caption{Characteristic angular frequencies, Eqs.~\eq{eq:Omega:ch} and~\eq{eq:Omega:ch:real}, time periods~\eq{eq:time:ch:real}, the continuity parameter $\Omega_{\ch} R$ and the enhancement factor~\eq{eq:f} for various strengths of the field and sizes of the circle.}
\label{tbl:examples:1}
\end{table*}

The integer number $M_{\Omega,B}$, the characteristic frequency $\Omega_{\mathrm{ch}}$ and the corresponding characteristic time period $\tau_{\mathrm{ch}}$ of the device can be written in physical units as follows:
\beqn
M_{\Omega,B} & \simeq & \left\lfloor \frac{5.1 \times 10^{6} \cdot {\bigl( \frac{\Omega}{1/\text{s}} \bigr)} {\left( \frac{B}{\text{T}} \right)} {\left( \frac{R}{\text{m}} \right)}^{3}}{
1 - 1.1 \times 10^{-17} \cdot {\bigl( \frac{\Omega}{1/\text{s}} \bigr)}^{2} {\left( \frac{R}{\text{m}} \right)}^{2}} \right\rfloor \,, 
\label{eq:M:ceiling:phys} \\
\Omega_{\mathrm{ch}} (B) & \simeq & 2 \times 10^{-7} \cdot {\left( \frac{B}{1 \, {\mathrm{T}}} \right)}^{-1} {\left( \frac{R}{1 \,{\mathrm{m}}} \right)}^{-3} \, {\mathrm{s}}^{-1}\, ,
\label{eq:Omega:ch:real} \\
\tau_{\mathrm{ch}} (B) & \simeq & \frac{2 \pi}{\Omega_{\ch}(B)} = 5.1 \times 10^{6} \cdot {\left( \frac{B}{1 \, {\mathrm{T}}} \right)} {\left( \frac{R}{1 \,{\mathrm{cm}}} \right)}^{3}   \, {\mathrm{s}}, \qquad
\label{eq:time:ch:real}
\eeqn
respectively. Values of the characteristic frequencies and time periods for a certain set of $B$ and $R$ are shown in Table~\ref{tbl:examples:1}.

In Fig.~\ref{fig:bands}  we illustrate the  structure of the enhancement bands for a circle of the radius $R=1$~cm. In the zeroth ($M_{\Omega,B}=0$) band -- filled by the reddish color in Fig.~\ref{fig:bands} -- the enhancement is absent and the zero--point energy is given by the $B=0$ expression, Eq.~\eq{eq:ZP:B0}, multiplied by two due to the presence of two scalar degrees of freedom in the complex scalar field. In the next band with $M_{\Omega,B}=1$ the enhancement of the rotational zero--point energy is equal to 13, while the highest shown band with $M_{\Omega,B} = 9$ gives the enhancement factor of 433. Thus, the magnetic flux enhances the negative--valued moment of inertia of the zero--point fluctuations. 

It is important to mention that the enhancement prefactor in the expression for the zero-point energy~\eq{eq:E:rotating:B} depends on the quantity $M_{\Omega,B}$ while this integer quantity is a stepwise function of the product $B \Omega$ (multiplied by a nonzero relativistic factor), Eq.~\eq{eq:M:ceiling}. Thus, the enhancement effect works only for a rotating device {\it and} only  in a background of the magnetic field (otherwise the product $B \Omega$ is zero). For a static device and/or in the absence of the magnetic field the enhancement effect is absent.

According to Eqs.~\eq{eq:E:rotating:B} and \eq{eq:M:ceiling:2}, one can distinguish three different limits in terms of the strength of the magnetic field $B$ and the angular frequency $\Omega$:
\begin{itemize} 
\item[(i)] In the ``continuous'' limit, 
\beqn
\Omega \gg \Omega_{\ch}\,, \qquad \mbox{or} \qquad e B \Omega R^{3} \gg 1\,,
\eeqn
the values of the integer number $M_{\Omega,B}$ become so large that $M_{\Omega,B}$ may be regarded as a continuous quantity:
\beqn
M_{\Omega,B} \bl_{\Omega \gg \Omega_{\ch} (B)} \simeq  \frac{\Omega}{\Omega_{\ch}} \frac{1}{1 - \Omega^{2} R^{2}}. \qquad
\label{eq:MF:nonrel}
\eeqn

\item[(ii)] In the nonrelativistic limit, 
\beqn
\Omega R \ll 1\,,
\eeqn
the denominator in Eqs.~\eq{eq:M:ceiling:2} and \eq{eq:MF:nonrel} can be neglected safely.

\item[(iii)] In the ``magnetic'' limit 
\beqn
\Omega_{\ch} R \ll 1\,, \quad \mbox{or} \quad e B \Omega R^{2} \gg 1\,,
\eeqn
the magnetically enhanced part provides a dominant contribution to the energy density~\eq{eq:E:rotating:B}.

\end{itemize}

For simplicity, let us simultaneously impose the nonrelativistic,\footnote{We usually consider nonrelativistic rotation $\Omega R \ll 1$, which is more suitable for an experimental setup as we will see below.} continuous and magnetic limits (these limits are consistent with each other):
\beqn
1 \gg \Omega R \gg \Omega_{\ch} (B) R\,.
\eeqn
Physically, these conditions correspond to slow rotation of the device in a strong magnetic field. In this limit the negative--valued rotational energy of the zero--point fluctuations 
grows quadratically with the strength of the magnetic field $B$:
\beqn
E^{\mathrm{ZP}}_{\Omega,B} \bl_{\Omega \gg \Omega_{\ch}(B)} = - \frac{R \Omega^{2}}{4 (\Omega_{\ch} R)^{2}} \equiv - \frac{e^{2} B^{2} R^{5}}{4} \Omega^{2} \,, 
\label{eq:E:rot:B}
\eeqn
and, consequently, the magnetic field enhances the negative moment of inertia of the zero--point fluctuations:
\beqn
I^{\mathrm{ZP}} = - \frac{e^{2} B^{2} R^{5}}{2}\,.
\eeqn

Notice that in the absence of the magnetic field the (negative--valued) rotational energy of the zero--point fluctuations is proportional the circle's radius~\eq{eq:ZP:B0}, while in the presence of the strong magnetic field the rotational energy grows as a fifth power of the radius~\eq{eq:E:rot:B}. 

The magnetic enhancement factor,
\beqn
f (B) & \equiv & \frac{E^{\mathrm{ZP}}_{\Omega,B}}{E^{\mathrm{ZP}}_{\Omega,B=0}}  \bl_{\Omega \gg \Omega_{\ch}(B)} \!\!\!\! = \frac{6}{(\Omega_{\ch} R)^{2}} \equiv 6 \, e^{2} B^{2} R^{4},
\qquad
\label{eq:f}
\eeqn
grows rapidly as the circle's radius and/or the strength of the magnetic field increases. According to Table~\ref{tbl:examples:1}, the enhancement factor $f$ may become an {\it astronomically} large quantity ($f \sim 10^{20}$ and higher) for macroscopically large objects in the presence of a strong, but experimentally feasible, magnetic field.

\subsubsection{Illustration of the enhancement due to magnetic field}

\begin{figure}[!thb]
\begin{center}
\includegraphics[scale=0.35,clip=false]{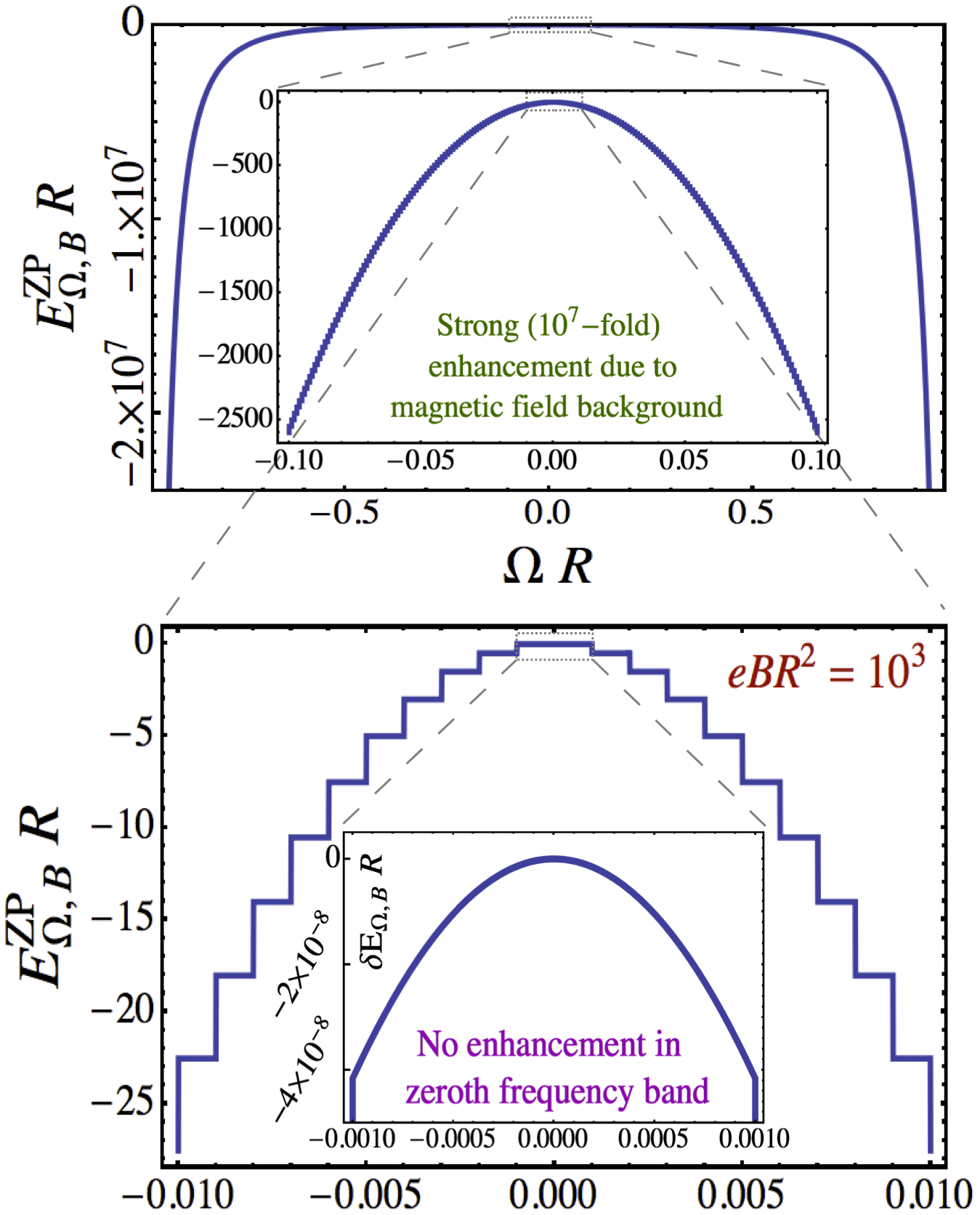}
\end{center}
\vskip -5mm
\caption{Rotational energy of zero--point fluctuations~\eq{eq:E:rotating:B} as a function of the angular frequency $\Omega$ at the background of the magnetic field $e B R^{2} =10^{3}$ (500 elementary fluxes pass through the circle) corresponding to $\Omega_{\ch} R = 10^{-3}$. Four different scales are shown (the description is given in the text).}
\label{fig:instability:strong}
\end{figure}

In the presence of the magnetic field, the rotational zero--point energy $E^{\mathrm{ZP}}_{\Omega,B}$, Eq.~\eq{eq:E:rotating:B}, becomes a nontrivial function of the angular frequency $\Omega$. A corresponding illustration for a relatively large flux of the magnetic field (500 elementary fluxes) is given in Fig.~\ref{fig:instability:strong}. Four different scales of the angular frequency $\Omega$ are shown. 

The upper plot in Fig.~\ref{fig:instability:strong} shows the energy at the whole range of frequencies, $ - 1 < \Omega R < 1$. The energy decreases unboundedly as the angular frequency increases. As we have already mentioned, the relativistic deep minimum of the system at $|\Omega| \to 1/R$ is an artifact which appears due to our assumption that the thickness of the circle is infinitely (mathematically) small. In physical spatially extended systems -- such as a cylinder -- the rotational zero--point energy has its minimum at finite values of the angular frequency~\cite{ref:I}.

The inset of the upper plot of Fig.~\ref{fig:instability:strong} shows the behavior of the energy in one-tenth of the whole range of frequencies, $ - 10^{-1}< \Omega R < 10^{-1}$. The energy dependence on frequency is an upside-down parabola with the maximum at the stationary point $\Omega = 0$. 

The lower plot of Fig.~\ref{fig:instability:strong} shows the energy at one order smaller frequencies than shown in the previous plot (one-hundredth of the full scale, $ - 10^{-2}< \Omega R < 10^{-2}$). One can clearly see that the parabola of the previous plot is, in fact, a discontinuous steplike function of the frequency. The discontinuities correspond to the boundaries between the bands which were already illustrated in Fig.~\ref{fig:bands}. Ten bands for each of the clockwise and counterclockwise directions are shown in the lower plot of Fig.~\ref{fig:instability:strong}.

Finally, the inset of the lower plot of Fig.~\ref{fig:instability:strong} shows the behavior of energy at an even smaller range of frequencies, one-thousandth of the full scale, $ - 10^{-3}< \Omega R < 10^{-3}$ (this range of frequencies corresponds to the wide zeroth band). Due to very weak dependence of the energy on the angular frequency, in this band we show the change in energy due to rotation, $\delta E^{\mathrm{ZP}}_{\Omega,B} = E^{\mathrm{ZP}}_{\Omega,B} - E^{\mathrm{ZP}}_{\Omega=0,B}$, rather than the energy itself. It turns out that the corresponding energy dependence on the angular frequency  -- which looked as flat in the main part of the lower plot of Fig.~\ref{fig:instability:strong} -- is, in fact, a smooth parabola. 

The huge enhancement effect of the negative rotational energy of the zero--point fluctuations can clearly be seen by a comparison of the insets in the upper and lower plots of Fig.~\ref{fig:instability:strong}. The inset of the lower plot demonstrates that the negative moment of inertia of the nonrotating, $\Omega = 0$, state is tiny. The magnitude of this moment of inertia is similar in scale to the one caused by a neutral scalar field, Eq.~\eq{eq:ZP:B0}. The magnetic field enhances this negative moment of inertia by seven orders of magnitude (at the chosen value of the magnetic flux) as one can see in the inset of the upper plot. In fact, the enhancement factor $f$ can be even (much) larger for larger devices ($f \sim 10^{23}$ for a centimeter-sized device in a modest, 1~T strong, magnetic field; Table~\ref{tbl:examples:1}).

\begin{figure}[!thb]
\begin{center}
\includegraphics[scale=0.35,clip=false]{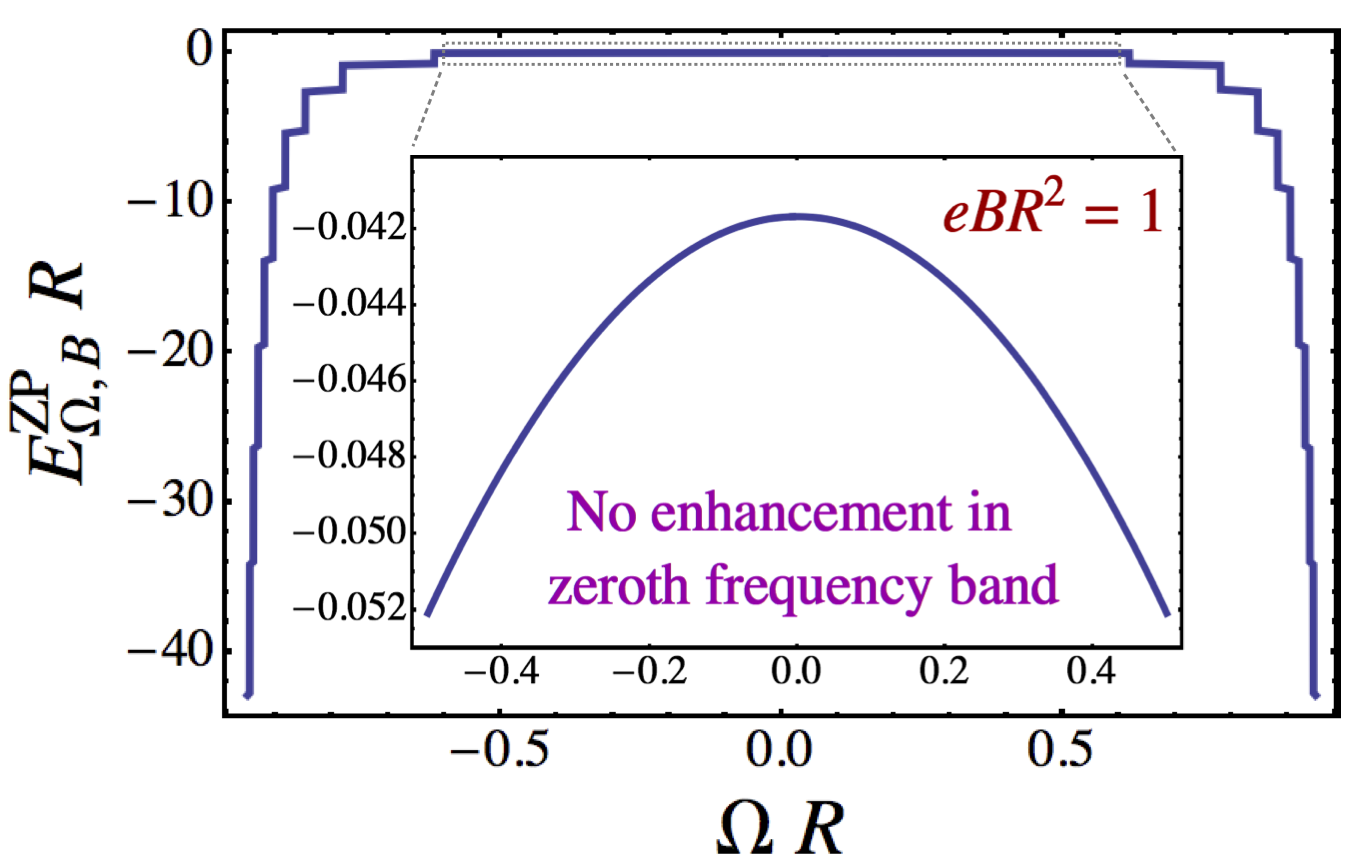}
\end{center}
\vskip -5mm
\caption{The same as in Fig.~\ref{fig:instability:strong} but for weaker magnetic field $e B R^{2} =1$ (one-half of the elementary magnetic flux passes through the circle, $\Omega_{\ch} R = 1$).}
\label{fig:instability:weak}
\end{figure}

The enhancement of the zero--point energy is much smaller at a weaker magnetic field. As an illustration, we show in Fig.~\ref{fig:instability:weak} the behavior of the energy of zero--point fluctuations for the magnetic field corresponding to one-half of elementary magnetic flux (i.e., the flux is thousand times smaller compared to the one of Fig.~\ref{fig:instability:strong}). One can see that the zeroth band widens drastically, while the energy dependence on the rotational frequency is still very weak. The strong energy minimum appears only when $\Omega$ approaches the relativistic limit,  $|\Omega| \to 1/R$. As we mentioned, this deep energy minimum is an artifact of the infinite thinness of our ``mathematical'' device and therefore the relativistic deep minimum is not present in physical, spatially extended systems.

\subsection{Rotational energy of massive devices}
\label{sec:rotation:massive}

So far we considered only the negative rotational energy of the zero--point fluctuations $E^{\mathrm{ZP}}_{\Omega,B}$ which turns out to favor a permanently rotating state. In a real physical case, the device itself should have a nonzero mass $m$ which should lead to nonzero positive classical rotational energy,
\beqn
E_{\mathrm{cl}}(\Omega) = \frac{I_{\mathrm{cl}} \Omega^{2}}{2} \equiv \pi \mu R^{3} \Omega^{2}\,,
\label{eq:E:cl}
\eeqn
where
\beqn
I_{\mathrm{cl}} \equiv \frac{\partial^{2} E_{\mathrm{cl}}}{\partial \Omega^{2}} = m R^{2}\,,  \qquad m = 2 \pi \mu R\,,
\eeqn
is the classical moment of inertia of the device and $\mu$ is the mass density per unit length of the device. As we will see below, the rotation in this case is nonrelativistic; therefore we are using the nonrelativistic formula for the classical energy~\eq{eq:E:cl}.

The classical energy favors a static state, $\Omega = 0$, so that a possible emergence of the permanently rotating state is conditioned by a competition between the quantum, zero-point part and the classical (mechanical) {\it rotational} part of the total energy:
\beqn
E(\Omega) = E^{\mathrm{ZP}}_{\Omega,B} + E_{\mathrm{cl}} (\Omega)\,.
\label{eq:E:total}
\eeqn
If the total energy $E(\Omega)$ has a {\it global} minimum at $\Omega \neq 0$ then the ground state corresponds to a permanent rotation.

Below we demonstrate the effect of the weakest possible magnetic enhancement in its most modest realization, corresponding to the first enhancement band of Fig.~\ref{fig:bands}. The higher bands, which correspond to stronger magnetic field and/or higher angular frequency, should give a much stronger effect, which will be considered elsewhere in application to a concrete physical device~\cite{ref:preparation}. 

\begin{figure}[!thb]
\begin{center}
\includegraphics[scale=0.35,clip=false]{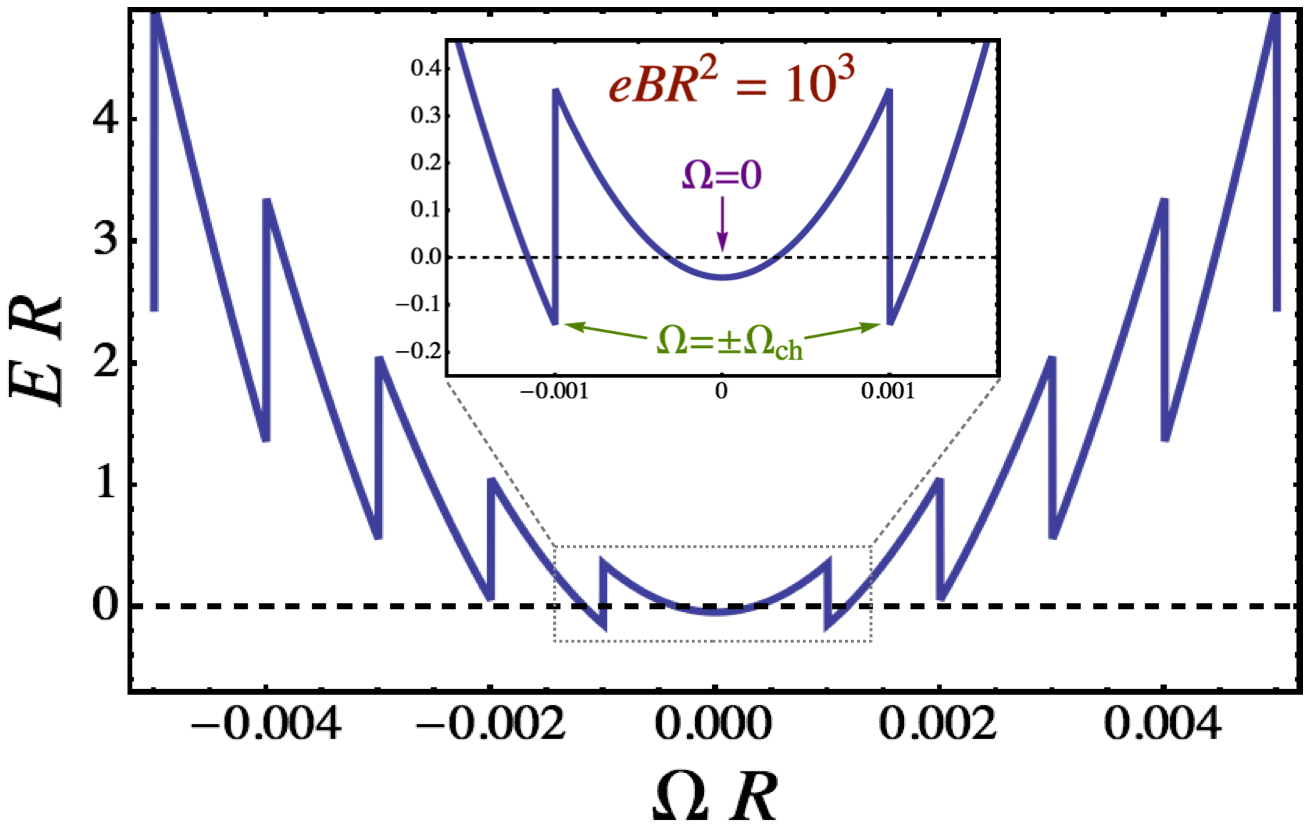}
\end{center}
\vskip -5mm
\caption{Illustration of the energetic favorability of a $\Omega {\neq} 0$ state: the total energy $E$, Eqs.~\eq{eq:E:rotating:B}, \eq{eq:E:total} and \eq{eq:E:cl},  as a function of the angular frequency~$\Omega$ for a device with the classical moment of inertia $I^{\mathrm{cl}} = 8\times 10^{5} R$ in the background of the magnetic field $e B R^{2} =10^{3}$.}
\label{fig:nontrivial:minimum}
\end{figure}

Figure~\ref{fig:nontrivial:minimum} shows a typical dependence of the total energy~\eq{eq:E:total} on the angular frequency $\Omega$. The inset illustrates the effect of the most modest magnetic  enhancement of the rotational zero--point energy, which is realized at the angular frequency $\Omega = \Omega_{\ch}$. At this angular frequency the classical (mechanical) part of the energy of the device is\footnote{Relativistic corrections are omitted in our considerations in this section since we consider the limit of nonrelativistic rotation, $\Omega_{\ch} R \ll 1$. The condition of the nonrelativistic rotation can also be formulated as a requirement for the magnetic flux to be much larger than one-half of the elementary flux: $F_{B} \equiv \pi R^{2} B \gg \pi / e$.}:
\beqn
E_{\mathrm{cl}} (\Omega_{\ch})  = \frac{I_{\mathrm{cl}} \Omega^{2}_{\ch}}{2} = \frac{\pi \mu}{e^{2} B^{2} R^{3}}\,,
\eeqn
while the zero--point part of the energy is given by Eq.~\eq{eq:E:rotating:B} with $M_{\Omega,B} = 1$:
\beqn
E^{\mathrm{ZP}}(\Omega_{\ch}) & \equiv & E_{\Omega_{\ch} + 0,B}  = - \frac{13}{24 R}\,, \qquad
\label{eq:E:q}
\eeqn
so that the total energy is
\beqn
E(\Omega_{\ch}) = \frac{\pi \mu}{e^{2} B^{2} R^{3}} - \frac{13}{24 R}\,.
\label{eq:E:Omega:ch}
\eeqn

At zero angular frequency the total energy~\eq{eq:E:total} is determined only by the zero--point part because the classical part of the total energy is vanishing
\beqn
E(\Omega=0) = E^{\mathrm{ZP}}_{\Omega=0,B} = - \frac{1}{24 R}\,.
\label{eq:E:Omega:0}
\eeqn

The permanently rotating state becomes a ground state if it has lower energy~\eq{eq:E:Omega:ch} compared to the energy~\eq{eq:E:Omega:0} of the static, $\Omega = 0$, device\footnote{In our article, we make a most conservative estimation of the effect and therefore we ignore a possibility for a ground state to be realized at higher frequencies, $\Omega = n \Omega_{\ch}$, $n \in \Z$ with $|n| \geqslant 2$.}:
\beqn
E(\Omega) < E(0) \,.
\label{eq:E:inequality}
\eeqn
This condition can also be written in the following form:
\beqn
 e^{2} B^{2} R^{2} > 2 \pi \mu\,,
\eeqn
which determines a minimal (critical) radius of the circle at fixed magnetic field $B$, or, equivalently, a minimal (critical) strength of the magnetic field at fixed radius of the device:
\beqn
R_{c}(B) = \frac{\sqrt{2 \pi \mu}}{e B}\,,
\label{eq:R:min}
\eeqn
such that for the circles of the radius $R > R_{c}(B)$ [or, equivalently in the background of the magnetic field $B > B_{c}(R)$ in the inverted Eq.~\eq{eq:R:min}] the lowest energy state corresponds to rotation with the frequency $\Omega = \Omega_{\ch}(B,R)$ [Eq.~\eq{eq:Omega:ch:hbar}], and ground state of the device corresponds to the permanent rotation.

Although the aim of this section is to demonstrate the theoretical existence of the new effect -- the enhancement of the rotational vacuum effect by the magnetic field --  it is still interesting to estimate the scales of the critical parameters~\eq{eq:R:min}. To this end, it is convenient to express the magnetic field and radius in Tesla and meters, so that Eq.~\eq{eq:R:min} can be rewritten as follows:
\beqn
R_{c} = 2.8 \times 10^{6} \cdot {\left( \frac{\mu}{{\mathrm{kg/m}}} \right)}^{1/2} {\left( \frac{B}{{\mathrm{T}}} \right)}^{-1} \cdot {\mathrm{m}}\,.
\label{eq:Bc:perpetuum:ph}
\eeqn

Now, let us make a very naive estimation. Suppose, for example, that we have a hypothetical material with massless charged excitations which has the density of, e.g., aluminum, $\rho_{\mathrm{Al}} = 2.7 \times 10^{3}\,\text{kg}/\text{m}^{3}$. Then for a circle made of  wire of the diameter $d = 1$~mm  one gets the corresponding mass per unit length $\mu_{\mathrm{Al}} = \pi \rho_{\mathrm{Al}} d^{2}/4 = 2.1 \times 10^{-3}\,\text{kg}/\text{m}$, and the right-hand side of~Eq.~\eq{eq:Bc:perpetuum:ph} evaluates to $1.3 \times 10^{5}$. In order to reach the critical point, one should have either a compact (1 meter wide) but astronomically strong uniform magnetic field with the strength of about $10^{6}$~T, or one should consider a huge circle of diameter of $130$~km pieced subjected to a uniform field of strength 1~T. This proposal -- based on a ``usual'' material -- is not realistic from an experimental point of view. Below we consider a device made of a carbon nanotube for which a perpetual motion may probably be realized.

\section{The rotational vacuum effect in a torus made of carbon nanotube}
\label{sec:nanotube}

A real material which does have massless charged excitations in its spectrum is a metallic (armchair) carbon nanotube. In fact, the carbon nanotubes act as genuine one--dimensional quantum wires~\cite{ref:wires} with a relativistic massless branch of the excitation spectrum~\cite{ref:graphene}. The electrically charged massless excitations are described by a Dirac equation. The excitations propagate with the Fermi velocity
\beqn
v_{F} \approx 8.1 \times 10^{5}\, \mathrm{m}/\mathrm{s} \approx \frac{c}{300}\,.
\label{eq:vF}
\eeqn

The mass per unit length of a typical, e.g., a (10,10) armchair carbon nanotube is~\cite{ref:Leonard}:
\beqn
\mu = 3.24 \times 10^{-15}\,\text{kg}/\text{m}\,.
\label{eq:mu}
\eeqn
The carbon nanotubes are very light, so that the device made of this material should have a small classical moment of inertia $I^{\mathrm{cl}}$ supporting a perpetual rotation via the rotational vacuum effect. 

In this section, we make a {\it rough} estimation the rotational vacuum effect for a device made of a metallic carbon nanotube. The experimental setup is similar to the one depicted in Fig.~\ref{fig:circle}: The role of the circle is played by a torus made of the nanotube while the role of the cut is played by a suitable chemical doping which substitutes certain carbon atoms by other ``foreign'' atoms.  The doping should provide a sufficiently wide and large potential barrier which should separate the charged excitations at both sides of the barrier and prevent their tunneling from one side of the doped region to the other side.

\begin{figure}[!thb]
\begin{center}
\includegraphics[scale=0.65,clip=false]{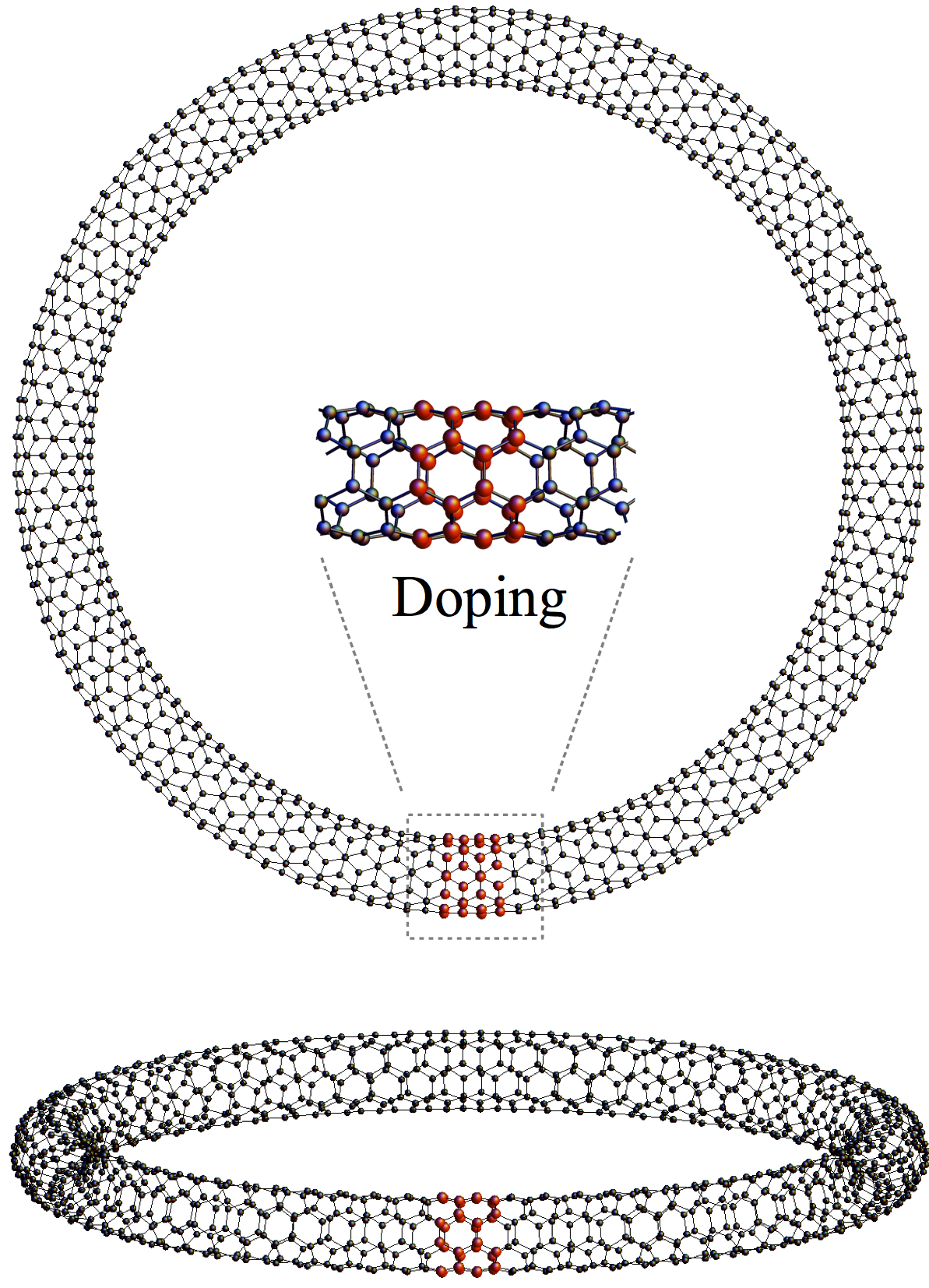}
\end{center}
\vskip -5mm
\caption{Suggested device made of a metallic carbon nanotube (the circle) with the doped region (the Dirichlet cut). }
\label{fig:doped:nanotorus}
\end{figure}

The aim of our calculation below is to estimate the typical scales (radii, angular frequency, and energy gaps) for the devices with the characteristic properties of ideal nanotubes in order to determine if in such systems the rotational vacuum effect is realizable in principle or not. To this end, it is enough to compute the enhanced rotational vacuum effect assuming the bosonic, and not fermionic, nature of the massless excitations, because in one spatial dimension the zero--point (Casimir) energies for a free massless scalar field and a free massless fermion field for certain boundary conditions are identical~\cite{Sundberg:2003tc}. For a carbon nanotube of a finite diameter the fermionic nature of the massless excitations should affect certain features of the zero--point energy~\cite{Bellucci:2009jr}. Indeed, the nanotube is a spatially two--dimensional system because it could be considered as a (two--dimensional) graphene sheet rolled into a cylinder. However, in our estimation we treat the nanotubes as very thin quantum wires~\cite{ref:wires}, because we study, basically, the long distance dynamics for sufficiently thin nanotubes. In our approximation the fermion excitations are treated as free particles so that we ignore the Coulomb interactions between the excitations.

Below, we consider a most modest realization of the magnetic enhancement in the first enhancement band with $M_{\Omega,B} = 1$ (see the illustration in Fig.~\ref{fig:bands}).

In order to adapt our formulae of the pervious section to the case of the nanotube torus we should 
\begin{itemize}
\item[(i)] notice that in carbon nanotubes, as in the graphene, the Fermi velocity $v_{F}$ plays the role of speed of light $c$ (so that we should make the substitution $c \to v_{F}$) and
\item[(ii)] take into account the double degeneracy of the mass\-less excitations in the carbon nanotubes (the zero--point energy should be multiplied by the factor of~2).
\end{itemize}

In our approximation the zero--point energy in a thin torus made of the doped nanotube is as follows
\beqn
E^{\mathrm{ZP}}_{\Omega,B} = - \Bigl[1 + 6  M_{\Omega,B} (M_{\Omega,B} + 1)\Bigr]  \frac{v_{F}^{2} + \Omega^{2} R^{2}}{12 R v_{F}} \hbar\,, \qquad
\label{eq:E:rotating:tube}
\eeqn
with
\beqn
M_{\Omega,B} = \left\lfloor \frac{e B \Omega R^{2}}{v^{2}_{F} - \Omega^{2} R^{2}} \, \frac{v_{F}}{\hbar} \right\rfloor \,.
\label{eq:M:ceiling:tube}
\eeqn 
The characteristic angular frequency is
\beqn
\Omega_{\ch} (B,R) = \frac{\hbar v_{F}}{e B R^{3}}\,.
\label{eq:Omega:ch:tube}
\eeqn

Following Section~\ref{sec:rotation:massive} we determine  the minimal critical radius of the torus which is required for the realization of the permanent rotation:
\beqn
R_{c}(B) = \frac{\sqrt{\pi \hbar \mu v_{F}}}{e B} \,.
\label{eq:R:min:tube}
\eeqn
For a torus of the radius $R > R_c(B)$ the lowest energy state corresponds to rotation with the angular frequency $\Omega = \Omega_{\ch}(B,R)$, Eq.~\eq{eq:Omega:ch:tube}.

By using Eqs.~\eq{eq:vF} and \eq{eq:mu}, the characteristic frequency~\eq{eq:Omega:ch:tube} and the critical radius~\eq{eq:R:min:tube} can be rewritten, respectively, as follows:
\beqn
\Omega_{\ch} \simeq 5.3 \times 10^{-10} \, {\left( \frac{B}{{\mathrm{T}}} \right)}^{-1} {\left( \frac{R}{{\mathrm{m}}} \right)}^{-3} {\mathrm{s}}^{-1},\quad
\label{eq:Omega:perpetuum:CNT}
\eeqn
and
\beqn
{\left( \frac{R_c}{{\mathrm{m}}} \right)} \simeq 0.0058 \, {\left( \frac{B}{{\mathrm{T}}} \right)}^{-1} \,.
\label{eq:Bc:perpetuum:CNT}
\eeqn

For the strongest static magnetic field achievable in laboratory conditions~\cite{ref:MagLab}, $B \approx 50$~T and for a modest magnetic field of the strength $B = 1$~T, the critical minimal radii~\eq{eq:Bc:perpetuum:CNT} are
\beqn
R_c({B = 50 \mathrm{T}})& \simeq & 1.2 \times 10^{-4}\, {\text{m}} \equiv 0.12 \,\mbox{mm}\,,
\label{eq:choice} \\
R_c({B = 1 \mathrm{T}}) & \simeq & 5.8 \times 10^{-3}\, {\text{m}} \equiv 5.8 \,\mbox{mm}\,. 
\eeqn
If $R > R_c$, then the negative zero--point energy of the nanotube torus wins over its classical energy~\eq{eq:E:inequality}, and the ground state of the torus should correspond to a permanent uniform rotation. The time period of the rotation at $R > R_c$ should be longer than the one given by the characteristic time $\tau_{\mathrm{ch}} \equiv 2 \pi / \Omega_{\ch}$ at $R = R_{c}$, respectively:
\beqn
\tau_{\mathrm{ch}}({B = 50\mathrm{T}})& \simeq &  0.92 \,\mbox{s}\,,
\label{eq:choice:time} \\
\tau_{\mathrm{ch}}({B = 1\mathrm{T}}) \simeq 2320\, {\text{s}} & \approx & 39 \,\mbox{min}\,. 
\eeqn

As the radius $R$ increases the period $\tau$ gets longer as the third power of the radius according to Eq.~\eq{eq:Omega:perpetuum:CNT}. 
Notice that Eqs.~\eq{eq:R:min:tube} and~\eq{eq:choice} were derived from Eq.~\eq{eq:M:ceiling:tube} by assuming a slow (nonrelativistic) rotation of the torus in its ground state. This assumption is well justified because $\Omega_{\ch} R_{\min} / v_{F} \sim 10^{-11} \dots 10^{-9}$ in our examples.

The nonrotating ($\Omega=0$) state is separated from the permanently rotating ($\Omega = \Omega_{\ch}$) state by the energy barrier $\delta E = \hbar v_{F}/R$. For the chosen set of parameters~\eq{eq:choice}, the energy barriers are as follows:
\beqn
\begin{array}{rcl}
\delta E (50 \mathrm{T}) & \simeq & 8 \times 10^{-25} \, \mathrm{J} \approx 4.6 \, \mu{\mathrm{eV}}\,,  \qquad \\
\delta E (1 \mathrm{T}) & \simeq & 1.4 \times 10^{-26} \, \mathrm{J} \approx 0.09 \, \mu{\mathrm{eV}}\,,
\end{array}
\label{eq:energy:gap1}
\eeqn
and the corresponding temperature scales, $T = \delta E/k_{B}$, are, respectively, as follows:
\beqn
T(50 \mathrm{T}) \simeq 0.053\,  \mbox{K}\,, \qquad
T(1 \mathrm{T}) \simeq 1 \times 10^{-3}\,  \mbox{K}. \qquad
\label{eq:energy:gap2}
\eeqn
At temperature $T \ll T_{c}$ the thermal transitions between different rotating states (and a static state) of the device should be rare. A slow cooling from $T$ to lower temperatures should allow for the device to exchange its angular momentum with the thermal bath and, eventually, to occupy its permanently rotating ground state which is favored energetically.

A fabrication of a carbon nanotube of this relatively large size~\eq{eq:choice}, supplemented with the doped region to emulate the Dirichlet cut, may be a technologically difficult task. However, the very aim of our estimation is to demonstrate that the strengths of the experimentally available magnetic field and the available materials may soon be suitable for fabrication of the permanently rotating devices which utilize the zero-point fluctuations. We suggest that an appropriate design of the device (using multiple cuts of possible specific profiles, multiwalled carbon nanotubes, etc.) and utilization of the magnetic enhancement of the rotational vacuum effect in higher bands will make it possible to diminish both the critical radius of the carbon nanotube and the time period of the rotation, and increase the temperature at which the device may function. 

\begin{figure}[!thb]
\begin{center}
\includegraphics[scale=0.09,angle=-90,clip=false]{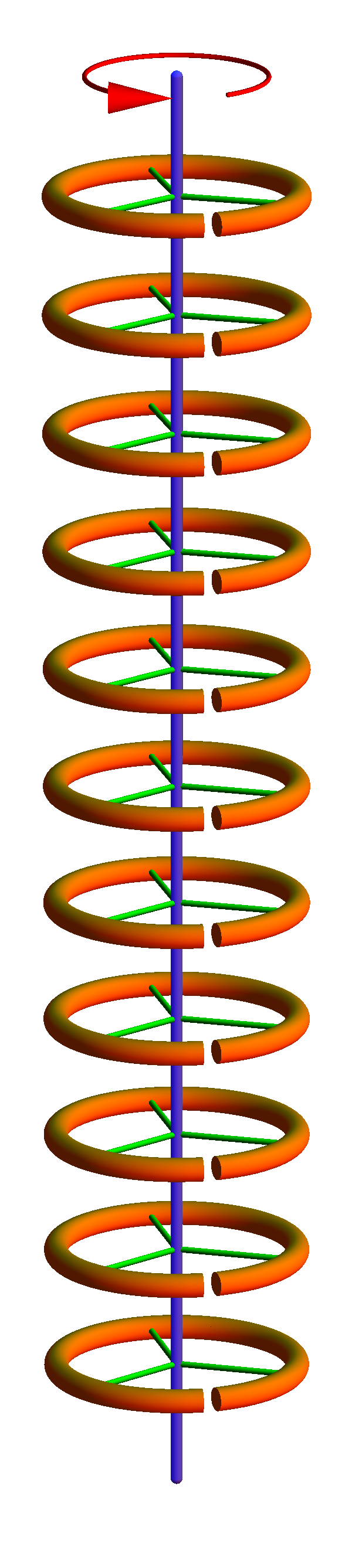}
\end{center}
\vskip -5mm
\caption{A simplest design of a macroscopic perpetuum mobile of the fourth kind made of a rigid array of elementary devices (the circles with the Dirichlet cuts, Fig.~\ref{fig:setup} or Fig.~\ref{fig:doped:nanotorus}). This design is visually similar to the very first proposal of a metamaterial~\cite{ref:split:ring} made of the C--shaped split-ring resonators.}
\label{fig:multidevice}
\end{figure}

There is another way to increase the operation temperature and, simultaneously, to avoid possible tunneling effects associated with the finite size of the device (Fig.~\ref{fig:setup} or Fig.~\ref{fig:doped:nanotorus}). We propose to assemble a large number $N$ of individual devices into a rigid periodic array along a certain axis as illustrated in Fig.~\ref{fig:multidevice}. Because of the additivity property, both the negative moment of inertia and the energy of this ``multidevice'' is $N$ times larger than, respectively, the moment of inertia and the associated energy of its individual component~\eq{eq:E:rotating:tube}. Thus, the characteristic frequency of the multidevice of Fig.~\ref{fig:multidevice} is the same as the characteristic frequency of its individual component~\eq{eq:Omega:ch:tube}, while the energy barriers between the rotating and nonrotating states [Eq.~\eq{eq:energy:gap1}] the associated temperatures~\eq{eq:energy:gap2}, should be $N$ times larger. The multidevice containing $N \sim 10^{5}$ elementary rings (i.e., circles with the Dirichlet cut), may operate at room temperature.

It is interesting to notice that the design of the perpetuum mobile of the fourth kind (Fig.~\ref{fig:multidevice}) is conceptually very similar to the simplest design of a metamaterial proposed first in Ref.~\cite{ref:split:ring} and experimentally confirmed later. The metamaterials are artificially engineered materials which have negative values for both permittivity $\varepsilon$ and permeability $\mu$ so that their refractive index is negative~\cite{ref:Veselago}. Both metamaterial and ``perpetuum mobile'' designs, Fig.~\ref{fig:multidevice}, use the same technological element, the C--shaped open ring (Fig.~\ref{fig:setup}). In the terminology of the metamaterial physics our elementary device  is referred as to the split--ring resonator. These resonators, placed along the direction of propagation\footnote{It coincides, in our case, with the axis of rotation in Fig.~\ref{fig:multidevice}.} of an electromagnetic wave constitute a metamaterial.

\section{Angular momentum and thermodynamics}
\label{sec:thermodynamics}

\subsection{Angular frequency of rotating bodies in thermal equilibrium}

In this section, we demonstrate that our idea of the  ``perpetuum mobile of the fourth kind driven by the zero--point fluctuations'' does not violate the laws of thermodynamics. 

The idea is obviously consistent with the first law of thermodynamics because no work is produced by the object which rotates in its ground (i.e., lowest energy) state. 

The second law of thermodynamics states that the entropy of any isolated system not in thermal equilibrium should increase and reach a maximum value in the equilibrium state. In our particular case, it is important to notice that a ground state of a {\it typical} macroscopically large rotating body should always correspond to zero angular frequency provided that this body interacts with an environment (for example, with a gas) via an exchange of angular momentum. Because of this interaction, the rotating body should lose a nonequilibrium part of its angular momentum, thus raising energy and entropy of the environment. The angular velocity of the body in its eventual thermal equilibrium should vanish: ${\boldsymbol \Omega} = 0$. 

Formally, one can prove the latter statement as follows. The angular velocity $\Omega$ is related to the energy of the system $E$ via the following relation\footnote{In Eq.~\eq{eq:Omega:E}, we use one-component quantities $\Omega$ and $L$, because we consider the rotation in a plane. In the three--dimensional space, they should be substituted by ${\boldsymbol \Omega} $ and ${\boldsymbol L}$, respectively.}~\cite{ref:LL5}:
\beqn
{\Omega} = {\left(\frac{\partial E}{\partial { L}}\right)}_{S}\,,
\label{eq:Omega:E}
\eeqn
where the angular momentum serves~$L$ as an independent extensive variable. In the vicinity of the thermal equilibrium the energy $E$ of a {\it typical} macroscopic system is always a smooth convex function of the angular momentum. Therefore the lowest energy state of this system should always correspond to $\Omega = 0$ since the derivative in Eq.~\eq{eq:Omega:E} should vanish at the energy minimum.

In a seemingly contradictory manner, in Ref.~\cite{ref:I} and in this paper we claim that there are certain objects which should be rotating permanently due to zero--point fluctuations, even if these objects are allowed to exchange angular momentum with an external environment such as a thermal bath. Below, we resolve this contradiction\footnote{The author sincerely thanks G.E.~Volovik for raising the question about consistency of perpetual rotation and thermodynamics.} by demonstrating that for systems driven by the rotational vacuum effect the thermodynamic relation~\eq{eq:Omega:E} is satisfied exactly despite the fact that the device rotates in its equilibrium state with a nonzero angular frequency, $\Omega \neq 0$.

\subsection{Angular momentum of zero--point fluctuations}

In general, the angular momentum $L^{\mu\nu}$ of a $d$--dimensional system can be expressed via a symmetric stress--energy tensor $T^{\mu\nu}$ as follows:
\beqn
L^{\mu\nu} & = & \int d^{d} x \, {\mathcal M}^{\mu\nu 0}(x)\,, \nonumber \\
{\mathcal M}^{\mu\nu\rho} & = & x^{\nu} T^{\mu\rho} - x^{\mu} T^{\nu\rho}\,.
\eeqn
Thus, the angular momentum of the zero--point fluctuations $L^{\mathrm{ZP}}_{\Omega,B}$ in our device can be expressed via the off-diagonal component $T^{\varphi 0}$ of the stress--energy tensor:
\beqn
L^{\mathrm{ZP}}_{\Omega,B} \equiv L_{z} \equiv L^{12} & = & R \int\limits_{0}^{2 \pi} d \varphi\, \left[ x^{2} \left\langle T^{10} \right\rangle -  x^{1} \left\langle T^{20}  \right\rangle \right] \nonumber \\
& = & R^{2} \int\limits_{0}^{2 \pi} d \varphi\, \left\langle T^{\varphi 0} \right\rangle \,.
\eeqn
According to Eq.~\eq{eq:Tmunu:B} the expectation value of this component is related to the Green's function as follows:
\beqn
\left\langle T^{\varphi 0}\right\rangle & = & \frac{i}{R} \,  \left[\frac{\partial}{\partial t} \left(\frac{\partial}{\partial \varphi'} + i \gamma_{B} \right) 
+  \left(\frac{\partial}{\partial \varphi} - i \gamma_{B} \right) \frac{\partial}{\partial t'} \right] \nonumber \\
& &  
G(t,t';\varphi,\varphi') \bl_{{}^{t' \to t}_{\varphi' \to \varphi}}\,. \qquad
\label{eq:Tphi0:B}
\eeqn

Using the explicit representation for the Green's function~\eq{eq:G:Omega:B:calc}, we get the following expressions for the density of the angular momentum of the zero--point fluctuations,
\beqn
{l}^{\mathrm{ZP}}_{\Omega,B} 
& = & - \Bigl[1 + 6 
M_{\Omega,B} (M_{\Omega,B} + 1)\Bigr]  \frac{\Omega}{24 \pi} \,, \qquad
\label{eq:Tphi0:rotating:B:dens}
\eeqn
and for the total angular momentum:
\beqn
{L}^{\mathrm{ZP}}_{\Omega,B} & \equiv & R \int\nolimits_{0}^{2\pi} d\varphi\, {l}^{\mathrm{ZP}}_{\Omega,B} \nonumber \\
& = & - \Bigl[1 + 6 
M_{\Omega,B} (M_{\Omega,B} + 1)\Bigr]  \frac{\Omega R}{12} \,, \qquad
\label{eq:Tphi0:rotating:B}
\eeqn
where the integer number $M_{\Omega,B}$, Eq.~\eq {eq:M:ceiling}, depends on the strength of the background magnetic field $B$ and on the angular frequency $\Omega$. Notice that the expectation value $\left\langle T^{\varphi 0}\right\rangle$ is a finite quantity and no time--splitting regularization is, in fact, needed.

\subsection{Relation between energy and angular momentum}

Surprisingly, the angular momentum~\eq{eq:Tphi0:rotating:B} and the energy~\eq{eq:E:rotating:B} of the zero--point fluctuations are related to each other by a ``classical'' relation:
\beqn
L^{\mathrm{ZP}}_{\Omega,B} \bl_{\frac{e B \Omega R^{2}}{1 - \Omega^{2} R^{2}} \notin \Z} = \frac{\partial E^{\mathrm{ZP}}_{\Omega,B}}{\partial \Omega}\,,
\label{eq:L:ZP}
\eeqn
provided that the angular frequency $\Omega$ at given magnetic field $B$ does not correspond to the discontinuities of both the energy and angular momentum of the zero--point fluctuations. 

One can check a self-consistency of our approach by using a slightly different derivation of the angular momentum of the zero point fluctuations~\eq{eq:Tphi0:rotating:B}. According to a general thermodynamic relation~\cite{ref:LL5}, the angular momentum $L$ of a rotating object is related to its energy ${\widetilde E}$ in the corotating frame~\eq{eq:coordinates},
\beqn
{\widetilde E}(L) = E(L) - \Omega \, L\,,
\label{eq:tilde:E}
\eeqn
as follows\footnote{Notice that despite Eqs.~\eq{eq:L:ZP} and~\eq{eq:rotating} differing only by a sign factor, these are different equations: The energy $E$ in Eq.~\eq{eq:L:ZP} is the energy of the rotating body in the laboratory (inertial) frame while the energy $\widetilde E$ in Eq.~\eq{eq:rotating} is the energy in the noninertial coordinate system rotating with the body.}:
\beqn
L= -  {\left(\frac{\partial {\widetilde E}}{\partial \Omega}\right)}_{S}\,.
\label{eq:rotating}
\eeqn
At vanishing temperature the entropy of the zero-point fluctuations is zero, so that one can neglect the fixed entropy condition in Eq.~\eq{eq:rotating}. 

The zero--point energy~\eq{eq:tilde:E} in the corotating reference frame can be calculated with the help of Eqs.~\eq{eq:E:rotating:B} and \eq{eq:Tphi0:rotating:B}:
\beqn
{\widetilde E}^{\mathrm{ZP}}_{\Omega,B} =  - \Bigl[1 + 6  M_{\Omega,B} (M_{\Omega,B} + 1)\Bigr]  \frac{1 - \Omega^{2} R^{2}}{24 R} \,. \qquad
\label{eq:tilde:E:rotating:B}
\eeqn
It is easy to check that the angular momentum of zero--point fluctuations~\eq{eq:tilde:E:rotating:B} and the corresponding energy in the corotating frame~\eq{eq:tilde:E:rotating:B} satisfy the thermodynamic relation~\eq{eq:rotating}.

In the absence of the magnetic field the energy of the  real-valued scalar field in the corotating reference frame is as follows:
\beqn
{\widetilde E}^{\mathrm{ZP}}_{\Omega} =  - \frac{1 - \Omega^{2} R^{2}}{48 R} \,. \qquad
\label{eq:tilde:E:rotating:B:real}
\eeqn

The zero-point energies of the charged~\eq{eq:tilde:E:rotating:B} and neutral~\eq{eq:tilde:E:rotating:B:real} scalar fields in the corotating (noninertial) frame have a different sign in front of the $\Omega^2$ term compared to the corresponding energies in the laboratory (inertial) frame, Eq.~\eq{eq:E:rotating:B} and Eq.~\eq{eq:ZP:B0}, respectively. We would like to stress that the experimentally measured energy of a rotating body is performed in the inertial laboratory frame and not in the noninertial corotating frame (we also would like to remind that the definition of the corotating frame depends explicitly on the angular frequency of the rotation of the object). Thus, it is the expressions~\eq{eq:ZP:B0} and \eq{eq:E:rotating:B} that determine the contributions of the zero-point energy to the energy balance of the neutral and charged systems, respectively.

\subsection{Perpetual motion: Thermodynamics}

The classical part of the angular momentum of our system is
\beqn
L^{\mathrm{cl}} = I^{\mathrm{cl}}  \Omega \equiv \frac{\partial E^{\mathrm{cl}}}{\partial \Omega} \qquad \mbox{with} \qquad
E^{\mathrm{cl}} = \frac{I^{\mathrm{cl}} \Omega^{2}}{2}\,, \quad
\label{eq:LE:cl}
\eeqn
where $I^{\mathrm{cl}}$ is the moment of inertia of the device.

The total energy (angular momentum) of the system is given by the sum of the energy (angular momentum) of the zero--point fluctuations 
and of circle itself\footnote{We neglect all other effects of the magnetic field on the energy of the system which are not essential for this discussion.}:
\beqn
E = E^{\mathrm{cl}} + E^{\mathrm{ZP}} \,,  \qquad  L = L^{\mathrm{cl}} + L^{\mathrm{ZP}}\,.
\label{eq:EL:total}
\eeqn
Equations~\eq{eq:E:rotating:B}, \eq{eq:Tphi0:rotating:B}, \eq{eq:LE:cl} and \eq{eq:EL:total} define a function $E = E(L)$ via the parametric dependence on the angular frequency $\Omega$.

\begin{figure}[!thb]
\begin{center}
\includegraphics[scale=0.27,clip=false]{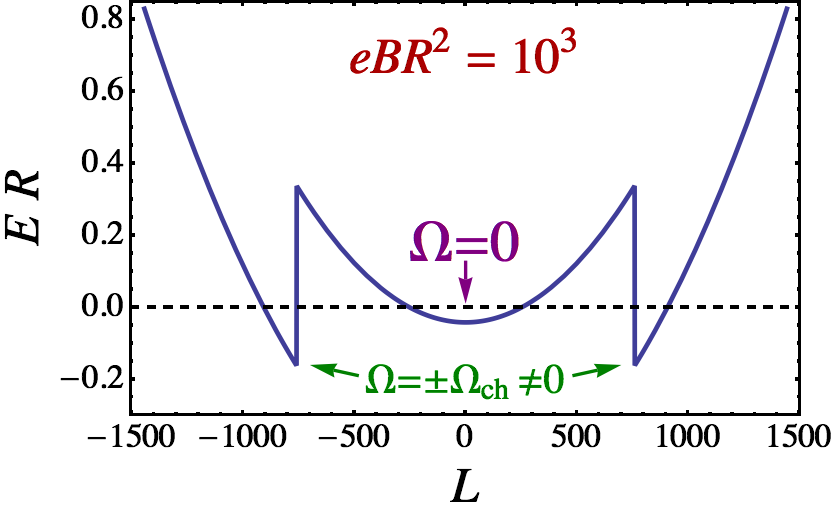}
\end{center}
\vskip -5mm
\caption{Total energy $E$ as a function of the total angular momentum $L$ calculated with the help of Eqs.~\eq{eq:E:rotating:B}, \eq{eq:Tphi0:rotating:B}, \eq{eq:LE:cl} and \eq{eq:EL:total} for a device with the classical moment of inertia $I^{\mathrm{cl}} = 8\times 10^{5} R$. A vicinity of $L = 0$ is shown only.
}
\label{fig:EvsL}
\end{figure}

In Fig.~\ref{fig:EvsL} we show an example of a typical behavior of the total energy of our device as a function its total angular momentum in a case when permanent rotation is favored. 
The ground state is doubly degenerate so that the system chooses the direction of rotation spontaneously.

It is important to notice that the energy $E = E(L)$ is not a regular smooth function of the angular momentum~$L$. Moreover, the discontinuities in the energy $E(L)$ appear precisely at those values of the angular momenta where the energy $E$ has its minima (with the exception for the standard local minimum with $L=0$).

The situation is very similar to a simple problem of finding a classical ground state of a particle in the following potential:
\beqn
V(x) = \left\{\begin{array}{ll} +\infty\,, \quad & x < 0\,, \\ x\,, & x \geqslant 0 \,. \end{array} \right.
\eeqn
The classical ground state $x = 0$ corresponds to the minimum of the potential $V(x)$, while, obviously, the first derivative of the potential with respect to $x$ cannot be computed at $x = 0$.
Moreover, the standard equation
\beqn
\frac{\partial V (x)}{\partial x} = 0
\eeqn
does not define the ground state, because $V(x)$ is not a smooth function of $x$.  Nevertheless, the ground state in this problem is well defined, while the derivative of $V$ can be computed as the following limit:
\beqn
\lim_{x \to + 0} \frac{\partial V (x)}{\partial x} = 1 \neq 0\,.
\eeqn

Coming back to thermodynamics of permanently rotating devices, we notice that the first derivative of the energy with respect to the angular momentum~\eq{eq:Omega:E} is not zero in the ground state $L=L_{\min} \pm 0$, where the choice of the sign should correspond to a nonsingular side of the energy minimum [i.e., a negative (positive) sign for the left (right) minimum in Fig.~\ref{fig:EvsL}]. Still, the ground state corresponds to a minimum of the energy.

Thus we come to the conclusion that the standard thermodynamical relation~\eq{eq:Omega:E} should be reformulated, due to the discontinuities, as follows:
\beqn
\Omega & = & \lim_{L \to L_{\min} \pm 0} {\left(\frac{\partial E}{\partial L}\right)}_{S} \,,
\eeqn
where the choice of the sign should correspond to a nonsingular side of the energy minimum. 

In summary, the discontinuous dependence of the rotational energy on the angular momentum due to zero--point fluctuations in the background of magnetic field guarantees the perpetual rotation of the device in its ground state $\Omega \neq 0$.

\section{Conclusions}

In Ref.~\cite{ref:I} it was shown that zero--point fluctuations may have a negative moment of inertia in a physical device with a very simple geometry. This leads to a counterintuitive effect that the absolute value of the {\it negative} rotational energy of the zero--point fluctuations increases with the increase of the angular frequency (the rotational vacuum effect). In the present paper we rederive the main result of  Ref.~\cite{ref:I} by using an explicit calculation via a Green's function approach via the time--splitting regularization. 

We have also shown that the presence of a magnetic field background may drastically enhance the negative moment of inertia of zero--point fluctuations so that at certain angular frequencies the negative rotational energy of the zero--point fluctuations may compensate the positive classical (mechanical) rotational  energy of the device. In this case the device becomes a perpetuum mobile of the fourth kind driven by the zero--point fluctuations which has the following surprising characteristics:

\begin{itemize}

\item[(i)]  the ground state of the device corresponds to a permanently rotating state;

\item[(ii)] the presence of an environment (for example, of a thermal bath) should generally not lead to a dissipation and to a cessation of rotation provided the ambient temperature is not too high;\\

\item[(iii)] the device has no internally moving parts (it is a mechanically rigid body).

\end{itemize}

We have also demonstrated that the very existence of this device is consistent with the laws of thermodynamics due to the absence of the energy transfer and due to specific discontinuities in the rotational energy of the zero--point vacuum fluctuations. 

As an illustration, we have roughly estimated the energy scales of a device made of a chemically doped, metallic carbon nanotube (Fig.~\ref{fig:doped:nanotorus}) and we have concluded that the zero--point energy of massless excitations in rotating torus-shaped doped carbon nanotubes may indeed overwhelm the classical energy of rotation for certain angular frequencies so that the permanently rotating state is energetically favored. A design of the macroscopically large, permanently rotating device at room temperature is proposed in Fig.~\ref{fig:multidevice}.

\acknowledgments 

The author is grateful to M.~Asorey, E.~Elizalde, K.~Kirsten, K.~Milton, M.~Plyushchay, D.~Vassilevich, and G.~E.~Volovik for interesting discussions and useful comments.  The author is thankful to M.~Schaden for valuable correspondence and important remarks.  The work was supported by Grant No. ANR-10-JCJC-0408 HYPERMAG (Agence nationale de la recherche, France).

\end{document}